\documentclass[twocolumn,superscriptaddress,showkeys,longbibliography,floatfix,
reprint,
nofootinbib,
 amsmath,amssymb,
 aps,
prb,
]{revtex4-2}
\usepackage[english]{babel}
\usepackage{xcolor}
\usepackage{graphicx}
\usepackage{dcolumn}
\usepackage{bm}
\usepackage{amsmath}
\usepackage{orcidlink}
\usepackage{braket}
\usepackage{natbib}
\usepackage[version=4]{mhchem}
\usepackage{gensymb}
\makeatletter
\let\old@makecaption=\@makecaption
\usepackage{subcaption}
\let\@makecaption=\old@makecaption
\makeatother
\usepackage{siunitx}
\usepackage{graphicx,multirow}
\usepackage{array}
\usepackage{hyperref}
\hypersetup{
    colorlinks,
    linkcolor={red!80!black},
    citecolor={blue!80!black},
    urlcolor={blue!80!black}
}
\usepackage{siunitx}
\usepackage[super]{nth}
\usepackage{lipsum}
\usepackage{orcidlink}
\usepackage{float}

\newcommand{\excop}{\hat{R}}

\newcommand{\beginsupplement}{%
        \setcounter{table}{0}
        \renewcommand{\thetable}{S\arabic{table}}%
        \setcounter{figure}{0}
        \renewcommand{\thefigure}{S\arabic{figure}}%
     }

\renewcommand{\selectlanguage}[1]{}
\begin{document}

\title{Dynamical Mean Field Theory for Real Materials on a Quantum Computer}

\author{Johannes Selisko\orcidlink{0000-0001-6733-0782}}
\affiliation{Corporate Sector Research and Advance Engineering, Robert Bosch GmbH, Robert-Bosch-Campus 1, D-71272 Renningen, Germany}

\author{Maximilian Amsler\orcidlink{0000-0001-8350-2476}}
\email{maximilian.amsler@de.bosch.com}
\affiliation{Corporate Sector Research and Advance Engineering, Robert Bosch GmbH, Robert-Bosch-Campus 1, D-71272 Renningen, Germany}

\author{Christopher Wever\orcidlink{0000-0003-4246-8464}}
\affiliation{Corporate Sector Research and Advance Engineering, Robert Bosch GmbH, Robert-Bosch-Campus 1, D-71272 Renningen, Germany}

\author{Yukio Kawashima\orcidlink{0000-0001-5918-1211}}
\affiliation{IBM Quantum, IBM Research Tokyo, 19-21, Nihonbashi Hakozaki-cho, Chuo-ku, Tokyo 103-8510, Japan}

\author{Georgy Samsonidze\orcidlink{0000-0002-3759-1794}}
\affiliation{Robert Bosch LLC, Research and Technology Center, Sunnyvale, CA 94085, USA}

\author{Rukhsan Ul Haq\orcidlink{0000-0002-6959-018X}}
\affiliation{IBM India Pvt. Ltd, 
IBM Research-India, 
Manyata Embassy Business Park, 
Outer Ring Road, 
Rachenahalli \& Nagawara Villages, 
Bengaluru, KA 560045, India}

\author{Francesco Tacchino\orcidlink{0000-0003-2008-5956}}
\affiliation{IBM Quantum, IBM Research Europe -- Z\"{u}rich, R\"{u}schlikon 8803, Switzerland}

\author{Ivano Tavernelli}
\affiliation{IBM Quantum, IBM Research Europe -- Z\"{u}rich, R\"{u}schlikon 8803, Switzerland}

\author{Thomas Eckl}
\email{thomas.eckl@de.bosch.com}
\affiliation{Corporate Sector Research and Advance Engineering, Robert Bosch GmbH, Robert-Bosch-Campus 1, D-71272 Renningen, Germany}

\date{\today}

\begin{abstract}
Quantum computers (QC) could harbor the potential to significantly advance materials simulations, particularly at the atomistic scale involving strongly correlated fermionic systems where an accurate description of quantum many-body effects scales unfavorably with size. While a full-scale treatment of condensed matter systems with currently available noisy quantum computers remains elusive, quantum embedding schemes like dynamical mean-field theory (DMFT) allow the mapping of an effective, reduced subspace Hamiltonian to available devices to improve the accuracy of \textit{ab initio} calculations such as density functional theory (DFT). Here, we report on the development of a hybrid quantum-classical DFT+DMFT simulation framework which relies on a quantum impurity solver based on the Lehmann representation of the impurity Green's function. Hardware experiments with up to 14 qubits on the IBM Quantum system are conducted, using advanced error mitigation methods and a novel calibration scheme for an improved zero-noise extrapolation to effectively reduce adverse effects from inherent noise on current quantum devices. We showcase the utility of our quantum DFT+DMFT workflow by assessing the correlation effects on the electronic structure of a real material, \ce{Ca2CuO2Cl2}, and by carefully benchmarking our quantum results with respect to exact reference solutions and experimental spectroscopy measurements. 

\end{abstract}


\maketitle




\section{Introduction}

Technological advances heavily rely on the design of innovative functional materials, a task chiefly driven by understanding and optimizing inherent relationships between processing, structure, and property. While AI-driven materials exploration are on the verge of gaining popularity and practical utility, numerical simulations at various length-scales are meanwhile routinely employed to support these efforts, with methods ranging from continuum and phase field modeling at the macro and meso-scale down to \textit{ab initio} approaches at the atomic level~\cite{louie_discovering_2021,marzari_electronic-structure_2021,ament_autonomous_2021}. For the latter, density functional theory (DFT) has emerged as the most popular technique due to its convenient accuracy at moderate computational cost~\cite{lejaeghere_reproducibility_2016}, despite the limitations arising from 
the use of approximated exchange-correlation functionals. 
For strongly correlated materials, however, semi-local DFT fails to correctly capture the underlying physics, thus calling for methods reaching beyond the mean field treatment of the electron interactions~\cite{simons_collaboration_on_the_many-electron_problem_direct_2020}.

An increasingly prevalent approach is to tackle such strongly correlated systems by extending DFT with an embedding scheme where a small subspace of correlated orbitals is treated with a many-body method~\cite{sun_quantum_2016}. Dynamical mean field theory (DMFT) is a common choice~\cite{georges_hubbard_1992}, which is based on describing correlated orbitals within a lattice as impurities embedded in a self-consistent, time-dependent mean field, and on the assumption that the lattice self-energy is local. Constructing DMFT orbitals from DFT for the subspace of partially occupied $d$ and $f$ shells in transition metal and rare earth elements, respectively, significantly improves the description of the electronic structures of strongly correlated condensed matter systems~\cite{kotliar_electronic_2006, sangiovanni_static_2006}. Such DFT+DMFT calculations have been meanwhile applied to describe the physics of a wide range of materials, including superconducting cuprates~\cite{karp_many-body_2020, karp_superconductivity_2022}, nickelates~\cite{karp_many-body_2020, karp_superconductivity_2022, chen_dynamical_2022}, and other Perovskite-type materials~\cite{yang_dynamical_2010,paul_strain_2019,park_srnbo3_2020,paul_cation_2020,si_pitfalls_2021,cappelli_electronic_2022}.

The computational complexity of DMFT itself depends on the method employed to solve the underlying Anderson impurity model (AIM). Exact diagonalization methods are only applicable to small systems~\cite{sangiovanni_static_2006,lu_exact_2017}, since the Hamiltonian matrix scales exponentially with the number of orbitals. Alternatively, an exact solution, within statistical errors, can be computed using quantum Monte Carlo (QMC) methods by expressing the impurity problem in a Lagrangian formulation in imaginary time~\cite{hirsch_monte_1986}. The limitations of its most popular flavor, the continuous time hybridization-expansion QMC (CTHYB)~\cite{werner_continuous-time_2006,gull_continuous-time_2011,seth_triqscthyb_2016}, arise from the potential Monte-Carlo sign problem and the possibly slow convergence particularly in the limit of low temperatures~\cite{loh_sign_1990}. Further, to obtain any physically meaningful quantities, the measured Green's function has to be analytically continued to the real frequency axis, a task that can be tedious and is, in general, ill conditioned~\cite{silver_maximum-entropy_1990}. In addition, approximate impurity solvers have been developed, such as Hubbard I~\cite{dai_calculated_2003,qiu_improved_2018}, density matrix~\cite{garcia_dynamical_2004} or numerical renormalization group~\cite{zitko_energy_2009}, which often trade better scaling for accuracy.

Quantum computers (QC) offer the potential to significantly improve upon the above mentioned classical impurity solvers~\cite{bauer_hybrid_2016,endo_calculation_2020,rungger_dynamical_2020,keen_quantum-classical_2020,jaderberg_minimum_2020} by providing an algorithm with advantageous, i.e., polynomial scaling or by improving the accuracy. The core advantage of quantum computing impurity solvers is achieved by mapping each fermionic spin orbital to a qubit, thus reducing the exponential complexity to a Hilbert space that can be expressed with a linearly increasing number of qubits. Several avenues to obtain the AIM Green's function have been proposed in the literature, the majority of which can be categorized 
as based on either the Lehmann representation~\cite{rungger_dynamical_2020,Rizzo_2022,  ehrlich_perspectives_2023}, a subspace expansion~\cite{jamet_quantum_2022}, or time-evolution~\cite{endo_calculation_2020,keen_quantum-classical_2020,libbi_effective_2022}, all with their respective advantages and disadvantages. For instance, time-evolution in fermionic systems leads to long circuits with large numbers of two-qubit operations, a challenge for current noisy quantum computers which only produce accurate impurity Green's functions for two-site toy models~\cite{keen_quantum-classical_2020}. On the other hand, the Lehmann representation requires, in addition to the knowledge of the ground state (GS), the calculation of electron and hole excited eigenstates, a task that demands an appropriate algorithm to retain favorable scaling for any applications beyond simple toy systems~\cite{ehrlich_perspectives_2023}. Common noisy quantum computing GS algorithms involve variational approaches~\cite{peruzzo_variational_2014, tilly_variational_2021, cerezo_variational_2021}, while excited states can be computed either with penalty function methods~\cite{penalty_method_Kuroiwa, penalty_method_Higgott}, subspace search~\cite{nakanishi2019subspace}, or the quantum equation of motion~\cite{Ollitrault_2020}.

In this work, we present our results on solving the dynamical mean field theory for a real materials system on a quantum hardware using an automated hybrid quantum-classical atomistic modeling workflow. Based on the DFT+DMFT framework, we investigate the correlation effects in \ce{Ca2CuO2Cl2} (CCOC)~\cite{grande_uber_1977,yamada_single-layer_2005,baptiste_ca2cuo2cl2_2018}, a cuprate superconductor~\cite{hiroi_probable_1994,argyriou_structure_1995} exhibiting physical properties that are believed to arise from a single correlated $d$ band in the low-energy spectrum~\cite{zhang1988effective, lee2006doping}. We map the electronic structure of CCOC to an effective Hubbard Hamiltonian, and extract the Green's function of a single-site AIM with up to 6 bath sites based on the Lehmann representation. To this end, we employ the quantum equation of motion (qEOM) approach~\cite{Rizzo_2022,Ollitrault_2020} with truncated excitations to reduce computational cost resulting in excellent scaling, and a hierarchy of mitigation schemes to alleviate the inherent errors of the quantum device. In particular, we introduce a novel algorithm to effectively suppress the variance arising from stochastic errors in zero-noise extrapolation schemes with equivalent or even lower computational cost, thereby reducing the average squared deviation from the state vector results by at least a factor of two. Using this scheme, we present QC results that show excellent agreement with classical ED reference calculations, and are able to correctly reproduce the renormalized single-particle spectrum from experimental angle-resolved photoemission spectroscopy (ARPES) of CCOC.


\section{Methods}

\begin{figure*}
     \begin{subfigure}[b]{0.8\textwidth}
         \centering
         \includegraphics[height=6.2cm]{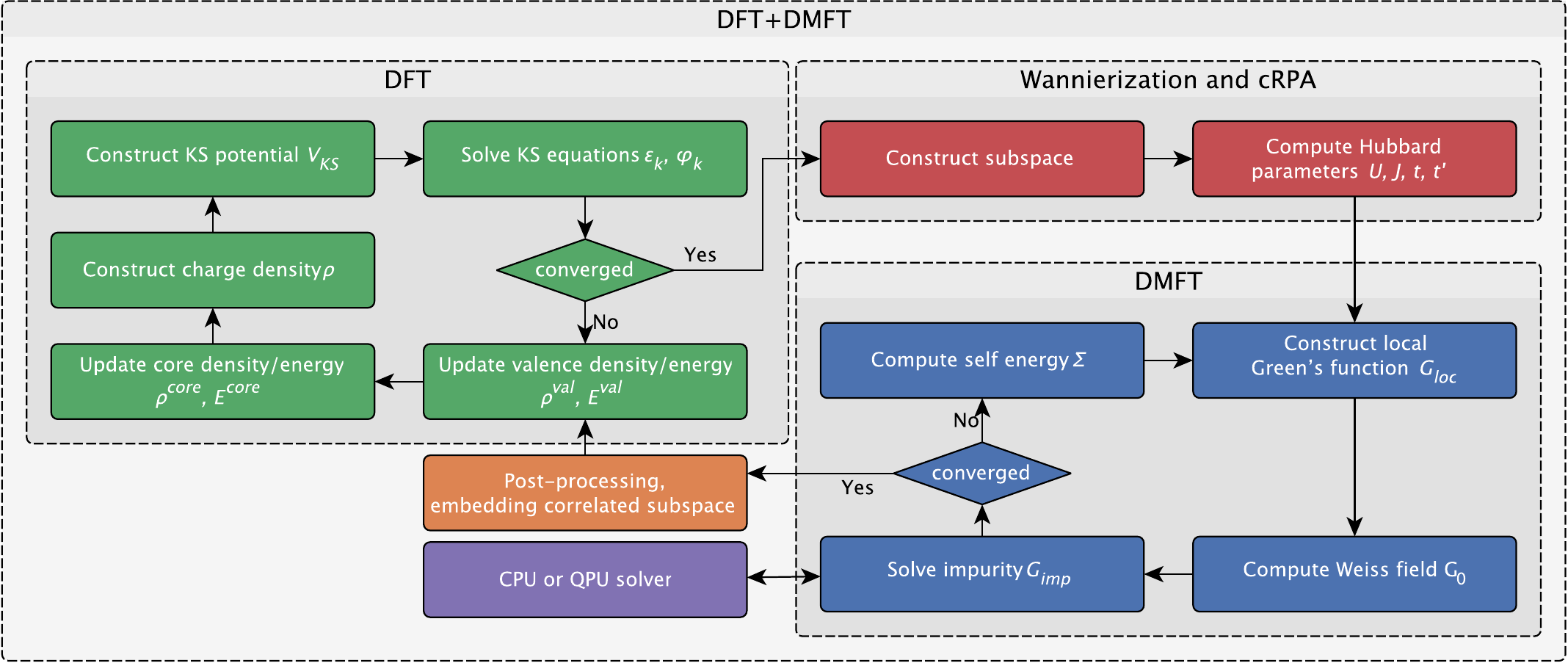}
         \caption{}
         \label{fig:flowcharta}
     \end{subfigure}
     \hfill
     \begin{subfigure}[b]{0.18\textwidth}
         \centering
         \includegraphics[height=6.2cm]{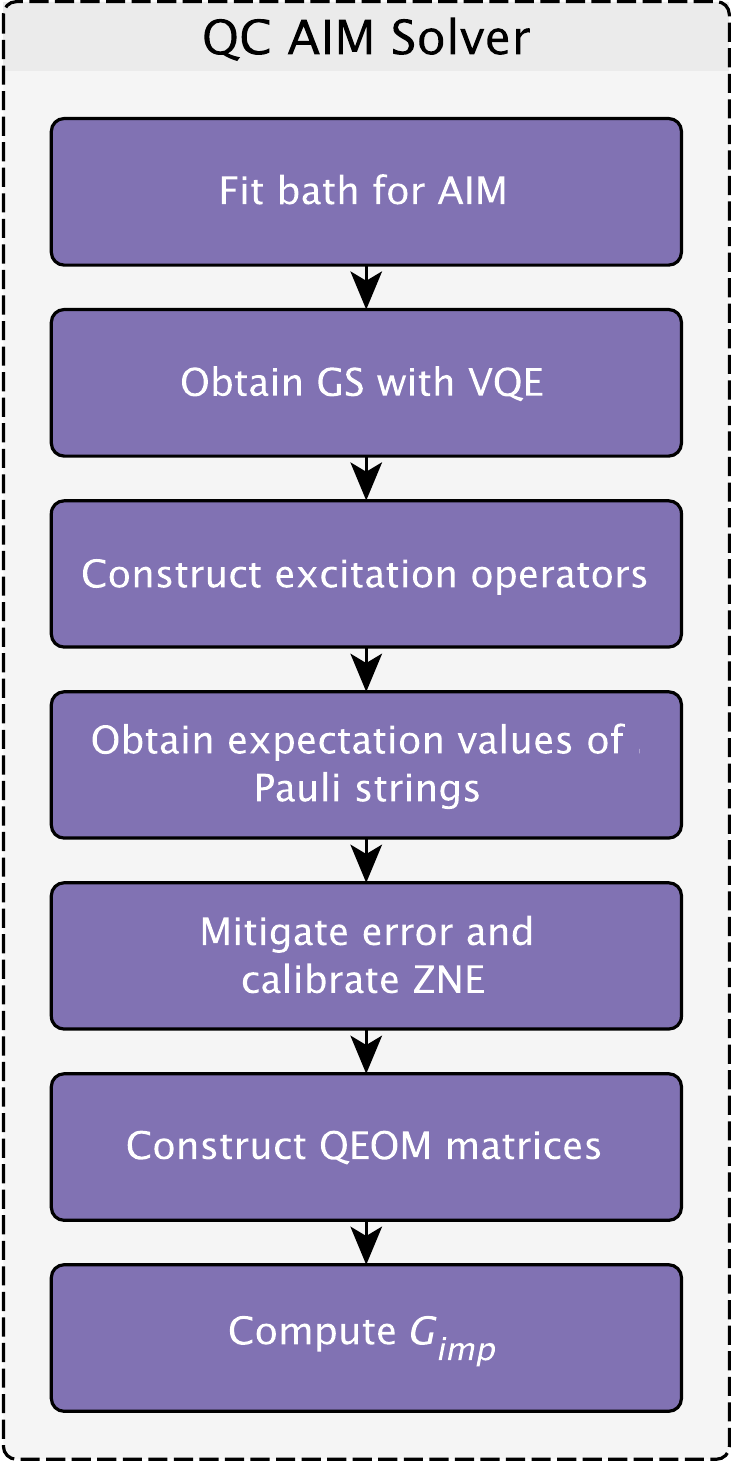}
         \caption{}
         \label{fig:flowchartb}
     \end{subfigure}
        \caption{Flowcharts illustrating the hybrid quantum-classical DFT+DMFT workflow. Subfigure (a) shows the overall flowchart, which starts out with solving the DFT self-consistency cycles as shown in green. The red blocks indicate the steps required to parametrize the subspace Hamiltonian by constructing localized Wannier orbitals and computing the interaction onsite and exchange parameters $U$ and $J$, respectively, as well as the hopping parameters $t, t'$, which serves as an input to the DMFT self-consistency cycle shown in blue. The impurity solver is denoted by the purple box, while the steps required for a charge-self-consistent DFT+DMFT or any post-processing steps are included in the orange block. The detailed flowchart of the QC AIM solver is shown in subfigure (b).}
        \label{fig:flowchartab}
\end{figure*}

We study CCOC given the structural parameters as determined experimentally in Ref.~\onlinecite{baptiste_ca2cuo2cl2_2018}, with Cu--Cu distances of 3.868~\AA\, within the \ce{CuO}-planes. The overall computational workflow to map this system to a problem that is solved within a hybrid quantum-classical DFT+DMFT cycle using a quantum impurity solver is shown in Fig.~\ref{fig:flowchartab}. In the following subsections, we describe in detail all components of the flowchart, thereby illustrating the steps involved in the relevant self-consistency cycles.

\subsection{DFT}
To construct the effective model Hamiltonian, we start out by computing the single-particle Bloch energy bands using density functional theory (DFT) as implemented in the Quantum ESPRESSO package~\cite{giannozzi_quantum_2009} which expands the wave function in a plane-wave basis. We employ the Perdew-Burke-Ernzerhof (PBE) approximation to the exchange-correlation functional~\cite{perdew_generalized_1996} and  norm-conserving pseudopotentials~\cite{van_setten_pseudodojo_2018}. The Kohn-Sham (KS) equations are solved self-consistently to obtain converged KS energies $\epsilon_k$  and orbitals $\varphi_k$, as shown in the DFT (green) block in Fig.~\ref{fig:flowcharta}. For CCOC, a plane-wave cutoff energy of 100~Ry together with a $12\times 12\times 12$  $k$-points mesh with a spacing smaller than 0.15/\AA\ result in total energies that are converged to within 5~meV/atom (see details in Sec.~\ref{sec:convergencedft} of the SI).

\subsection{Wannierization and cRPA}
The converged KS eigenstates are fed into 
a Wannierization framework to obtain localized orbitals, as shown in the red block in Fig.~\ref{fig:flowcharta}. We construct a low-energy model for the subsequent DMFT calculations by Wannierizing the single $d_{x^2-y^2}$ state crossing the Fermi level by using the Wannier90 package~\cite{pizzi_wannier90_2020}, with a large disentangling window of 18~eV and the lower and upper bounds at $-8$~eV and $10$~eV w.r.t. the Fermi energy, respectively. 

To obtain the effective, screened Coulomb interaction parameter $U$, we perform constrained random phase approximation (cRPA) calculations as implemented in RESPACK~\cite{nakamura-20-respack}. For this purpose, the Wannier orbital is converted with wan2respack~\cite{kurita_interface_2023} to a suitable format. For the cRPA calculation, we include 168 virtual orbitals in addition to the 32 occupied states, resulting in converged interaction parameters that we obtain from the static limit of the real part of the screened direct Coulomb integrals (see details in Sec.~\ref{sec:convergencecrpa} of the SI).

\subsection{DMFT}
The resulting, parametrized Hubbard model from the preceding step is then treated within the self-consistent DMFT Green's function formalism with a single impurity site, as illustrated by the blue block in Fig.~\ref{fig:flowcharta}. The DMFT calculations are performed with the software infrastructure provided by the Toolbox for Research on Interacting Quantum Systems (TRIQS)~\cite{triqs}.

In general terms, the DMFT self consistency cycle starts out by constructing the local Green's function $G_\text{loc}$ by integrating the lattice Green's function $G_\text{latt}$ over $k$, where 
\begin{equation}
    G_\text{latt}(\omega, k) = \frac{1}{\omega + \mu -\Sigma(\omega) + \epsilon(k)}
    \label{eq:glatt}
\end{equation}
with the chemical potential $\mu$ and the non-interacting single-particle DFT dispersion $\epsilon(k)$. We use the Dyson equation~\eqref{eq:dyson} to define a non-interacting Weiss field $G_0$, 
\begin{equation}
    \Sigma(\omega) = G_0^{-1}(\omega) - G_\text{imp}^{-1}(\omega).
    \label{eq:dyson}
\end{equation} 
An impurity solver is required to self-consistently compute the impurity Green's function $G_\text{imp}$, a task that can be either performed by a classical solver, or, as discussed later in Sec~\ref{sec:qcalgo}, with a quantum algorithm. After computing the self energy $\Sigma$ in Eq.~\eqref{eq:dyson}, the DMFT cycle repeats until convergence in $G_\text{imp}$ is reached.
The computationally expensive part of solving the impurity problem is shown in purple in Fig.~\ref{fig:flowcharta}, while the details of the QC-based solver is shown in Fig.~\ref{fig:flowchartb}. 

Initially, classical reference results are obtained using the CTHYB QMC impurity solver~\cite{seth_triqscthyb_2016} at a temperature of 386~K in a  paramagnetic setting. The total number of QMC cycles for each DMFT iteration is set to \num{1e8} with a cycle length of 400, resulting in an auto-correlation time of roughly 4. We perform a total of 25 DMFT iterations with the solid\_dmft workflow manager~\cite{Merkel2022,beck_charge_2022}, which conveniently incorporates DFT calculations with the DMFT toolbox of TRIQS. These pre-converged results subsequently serve as a starting point for self-consistent DMFT calculation with a discretized bath and an exact diagonalization (ED) solver at 0~K with the same frequency mesh as in CTHYB.

In principle, the solid\_dmft framework offers the possibility to feed a charge correction from the DMFT results back into the DFT cycle (e.g., in Quantum ESPRESSO) within an upfolding scheme, allowing a fully charge-self-consistent DFT+DMFT loop (shown by the orange box in Fig.~\ref{fig:flowcharta}). In this work, however, we terminate the cycle after one full DFT+DMFT step, a procedure commonly referred to as one-shot (OS) DFT+DMFT. In many cases, OS DFT+DMFT alone already offers a good description of the many-body effects governing the physics of strongly correlated materials~\cite{beck_charge_2022}.

\subsection{Mapping to Anderson Impurity Hamiltonian}
The following sections outline the methods involved to prepare and execute the quantum impurity solver within the DMFT cycle. The sequence of steps are illustrated in Fig.~\ref{fig:flowchartb}, which will be referenced throughout.

\subsubsection{Bath Discretization\label{sec:bath}} 
\begin{figure*}
 \centering
 \includegraphics[width=0.8\textwidth]{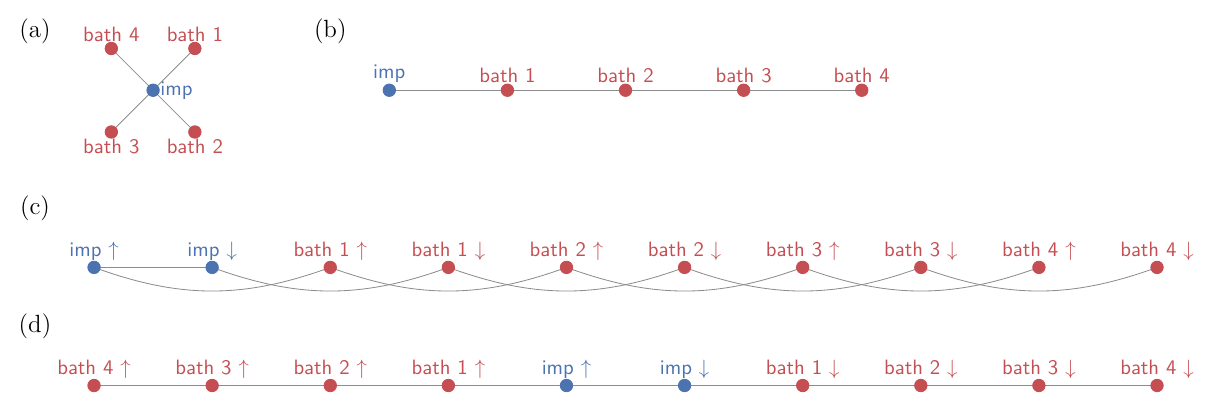}
\caption{An AIM for 4 bath sites in subfigure (a) for the star topology and in (b) for the chain topology. Subfigure (c) shows the Jordan-Wigner mapping of spin-orbitals to qubits, where the order is based on qubit indexing, and subfigure (d) illustrates the same mapping with re-indexed qubits. The grey lines in (c) and (d) connect sites which are coupled in the Hamiltonian.}
\label{fig:aim}
\end{figure*}

In contrast to QMC impurity solvers, methods based on exact diagonalization or quantum computing work in a Hamiltonian formulation of the impurity problem and require a mapping to an AIM with a finite number of bath sites, i.e., qubits. In our algorithm we perform this mapping (first block in Fig.~\ref{fig:flowchartb}) by fitting a discrete, non-interacting model with $N_b$ bath sites to the hybridization function describing the impurity problem:
\begin{equation}
    \Delta(i\omega_n) = i\omega_n-G_0^{-1}(i\omega_n)-\epsilon_0.
\end{equation}
Here, $G_0$ is the non-interacting impurity Green's function of our submodel and $\epsilon_0$ is the effective atomic energy level obtained from the Wannier orbital. The functions depend on the discrete, imaginary, fermionic Matsubara frequencies $i\omega_n$.
We use the discretization function from the TRIQS toolbox~\cite{triqs}, which fits the parameters of an impurity problem with a bath where each site is directly connected to the impurity site but not to other bath sites (star topology, denoted by a bar, see Fig.~\ref{fig:aim}). The hybridization of such a model is given by
\begin{equation}
    \Delta_\text{disc}(i\omega_n) = \sum_{j=1}^{N_b} \frac{\Bar{V_j}^2}{i\omega_n - \Bar{\epsilon_j}},
\end{equation}
where $\Bar{V_j}$ are the hopping strengths between the impurity and bath site $j$, and $\Bar{\epsilon_j}$ are the corresponding energy levels. We optimize the bath parameters with an additional $1/|\omega_n|$ weight to improve the fitting in the vicinity of the Fermi level, which is especially important for the DMFT convergence when $N_b<3$,
\begin{equation}
    \min_{\{\Bar{\epsilon_j}\}, \{\Bar{V_j}\}} \quad \sum_{\omega_n} \frac{|\Delta_\text{disc}(i\omega_n)-\Delta(i\omega_n)|}{|\omega_n|}.\label{eq:min_hyb_bath}
\end{equation}
For our AIM derived for CCOC, we carefully examine the convergence behavior with respect to the number of bath sites, and conclude that $N_b=6$ is more than sufficient to accurately reproduce the CTHYB hybridization function (see Sec.~\ref{sec:bath_sites} in the SI).

In order to obtain a model with suitable topology for current noisy quantum devices we perform a Lanczos tridiagonalization procedure, which produces an impurity model in a chain bath topology (see Fig.~\ref{fig:aim})~\cite{Lu_Efficient}. This model, with the impurity denoted with site index $j=0$, is described by the Hamiltonian
\begin{equation}
    \begin{split}
    \hat{H} = &\epsilon_0\hat{n}_0 + U\hat{n}_{0\uparrow}\hat{n}_{0\downarrow} + \\ &\sum_{j=1}^{N_b}\sum_{\sigma} \left[\epsilon_j\hat{n}_{j\sigma} + V_j\left(\hat{c}^\dagger_{j\sigma}\hat{c}_{j-1\sigma} + h.c.\right) \right], 
    \end{split}
    \label{eq:HAIM}
\end{equation}
and can be efficiently mapped to a linear chain of $2\times(N_b +1)$ qubits by enumerating the orbitals in the following order: $N_b\hspace{-1ex}\uparrow,\cdots, 1\hspace{-1ex}\uparrow, 0\hspace{-1ex}\uparrow, 0\hspace{-1ex}\downarrow, 1\hspace{-1ex}\downarrow, \cdots, N_b\hspace{-1ex}\downarrow$. In this notation $\hat{c}^\dagger_{j\sigma}$, $\hat{c}_{j\sigma}$, and $\hat{n}_{j\sigma}$ are the creation, annihilation, and number operators on site $j$ with spin $\sigma$, respectively, henceforth referred to as computational basis operators.
Using the Jordan-Wigner transformation~\cite{jordan_uber_1928}, the energy expectation value can be obtained by measuring the quantum circuit three times (independent of $N_b$), as all terms only involve a coupling of nearest neighbors. The sampling in the Pauli bases $XX\cdots XX$, $YY\cdots YY$ and $ZZ\cdots ZZ$ thus suffices to construct the expectation value of all terms in the Hamiltonian and subsequently obtain the energy.

\subsubsection{Exact Diagonalization} \label{sec:exact_diag}
Up to a small number of bath sites ($N_b\approx7$) one can obtain with reasonable numerical effort the zero-temperature Green's function by exact diagonalization using the Lehmann representation:
\begin{equation}
    G_\text{imp}(z) = \sum_{k>0} \frac{|\bra{k}\hat{c}^\dagger_0\ket{0}|^2}{z + (E_{0}-E_k)} + \frac{|\bra{k}\hat{c}_0\ket{0}|^2}{z - (E_{0}-E_k)}, \label{eq:lehm}
\end{equation}
where the sum runs over all energy eigenstates $\ket{k}$ with their associated energies $E_k$, and $k=0$ denotes the ground state. 
Depending on the purpose, we can choose the complex argument $z$ of the impurity Green's function in different ways: (i) we select the Matsubara frequencies $z=i\omega_n$ to continue DMFT iterations, or (ii) a real frequency with a small imaginary part $z=\omega +i0^+$ to obtain the final quantities of interest like the spectral function or the density of states (DOS). In contrast to the QMC methods operating on the imaginary time axis where the measured Green's function has to be analytically continued to the real frequency axis~\cite{silver_maximum-entropy_1990}, this process is unnecessary for the real frequency Green's functions obtained by the Lehmann representation, offering a significant advantage.

However, a challenge remains when the impurity Green's function $G_\text{imp}$ from a discrete spectrum is used to calculate the electronic self-energy in the Dyson equation on the real axis~\eqref{eq:dyson}.
The peaks in the inverse Weiss field $G_0^{-1}(\omega)$ need to be cancelled by peaks in $G_\text{imp}^{-1}(\omega)$, otherwise the imaginary part of the self-energy will be positive at some frequencies. This would lead to a negative, unphysical spectral function, which is given by $-1/\pi$ times the imaginary part of the lattice Green's function of Eq.~\eqref{eq:glatt}. Such spurious artifacts may arise not only due to the noise of current quantum devices, but already appear from the limited machine precision of classical computers.

To avoid this issue, we adapt the ideas of Lu~\textit{et al.} in Ref.~\onlinecite{Lu_Efficient} where the Green's function is stored as $N$ poles $(E_{k0}, \lambda_k)$, while $E_{k0}:=E_k-E_0$ is the energy difference in the denominator of Eq.~\eqref{eq:lehm} and $\lambda_k$ is the corresponding overlap in the numerator. The inverse of the Green's function can again be stored as a pole Green's function by diagonalizing an $N\times N$ matrix. In order to reduce the size of this matrix, one can eliminate some of the poles by distributing the overlap of small poles onto their neighbors. Ref.~\onlinecite{Lu_Efficient} describes a method to perform this procedure that locally preserves the \nth{0} and \nth{1}  moments, a method we also use in the evaluation of the Dyson equation in Eq.~\eqref{eq:dyson} to redistribute the peaks of $G_0^{-1}(\omega)$ with positive weights until only poles with negative weights remain. In our workflow, we use this pole-reduction purely as a post processing step to compute, e.g., the spectral function or the quasi particle weight (see orange box in Fig.~\ref{fig:flowcharta}).

\subsection{Quantum Algorithms\label{sec:qcalgo}}


\subsubsection{Ground State\label{sec:vqe}}

After constructing the AIM Hamiltonian as described in Sec.~\ref{sec:bath}, preparing an accurate GS is a necessary prerequisite for the quantum equation of motion method in Sec.~~\ref{sec:qeom} to compute the impurity GF in the Lehmann representation (second block in Fig.~\ref{fig:flowchartb}). In fact, the preparation of the GS of a many-body Hamiltonian $\hat{H}$ is a central challenge in accurately characterizing the electronic structure of any system. Here, we employ the variational quantum eigensolver (VQE) for this task~\cite{peruzzo_variational_2014, tilly_variational_2021, cerezo_variational_2021}. However, it is essential to highlight the versatility of our implementation, enabling the application of alternative approaches besides variational methods to construct the GS.

VQE is a hybrid quantum-classical algorithm designed to find an approximate solution to the Schrödinger equation for a given Hamiltonian $\hat{H}$. It utilizes a parameterized quantum circuit to prepare an ansatz wavefunction $| \psi(\boldsymbol{\theta}) \rangle$, where $\boldsymbol{\theta}$ is a set of variational parameters. These parameters are iteratively adjusted using classical optimization techniques to minimize the energy expectation value of the Hamiltonian $\langle \psi(\boldsymbol{\theta}) | \hat{H} | \psi(\boldsymbol{\theta}) \rangle$ estimated on a quantum device. The iterative refinement continues until convergence, at which point the ansatz with the final converged parameters provides an approximation to the GS of the system.

Challenges in applying VQE for GS preparation include the selection of a valid ansatz and the initialization of the circuit parameters for the classical optimizer, the overcoming of barren plateaus and local minima in the energy landscape, and the mitigation of inherent noise on near-term quantum devices. The choice of ansatz and the number of variational parameters significantly impact the accuracy of the GS approximation in VQE. Balancing ansatz expressiveness with computational resource requirements is essential for a successful application of the VQE. 


Our implementation utilizes a hardware-efficient ansatz with the linear entanglement structure~\cite{Kandala2017,ravi-22-cafqa} illustrated in Fig.~\ref{fig:ansatz} of the SI. This ansatz choice shows a good circuit expressibility for our Hamiltonians while featuring shallow circuit depths and supporting the limited qubit connectivity on IBM Quantum hardware. The ansatz circuit is complemented by an initial state circuit for preparing an initial state for the classical optimization. The initial state circuit can be either an empty circuit for preparing a zero initial state~\cite{onionvqe}, or a Slater determinant circuit~\cite{jiang2018slaterdeterminant} for preparing the mean-field (MF) state of the AIM Hamiltonian. The MF state is obtained by solving the AIM Hamiltonian in a generalized Hartree-Fock (GHF) approximation using the PySCF software package~\cite{sun_recent_2020}. The molecular orbital (MO) coefficients from the GHF solution are then used to parameterize the initial state circuit that prepares a Slater determinant following the method of Jiang~\textit{et al.}~\cite{jiang2018slaterdeterminant}.

The VQE optimization is performed using the limited-memory Broyden-Fletcher-Goldfarb-Shanno bound optimizer (L\_BFGS\_B)~\cite{Byrd_BFGS_1995,Zhu_BFGS_1997}, which we identify as the most robust method after comparing different approaches~\cite{tilly_variational_2021,bonet-monroig_performance_2023}, with finite difference gradients, employing \verb|Qiskit|'s default convergence criteria ($2.22 \times 10^{-15}$ as a relative tolerance for termination and 15000 as a maximum number of iterations). For variational parameter initialization, we adopt random initialization as well as several different implementations of the identity block method~\cite{Identity_Block_paper}. When combined with the GHF initial state, the identity block approach ensures that the optimization starts closer to the optimal point in the parameter space compared to random initialization. This, in turn, reduces the probability of encountering barren plateaus during optimization. Since the focus of this work is mainly on ultimately computing the impurity GF on a quantum computer, we do not perform the VQE calculations on the IBM Quantum hardware, but rather on a noiseless state vector (SV) simulator.


\subsubsection{Quantum Equation of Motion\label{sec:qeom}}
We employ the quantum equation of motion (qEOM) method to compute the excited states required for the Lehmann representation of the impurity Green's function. 
The qEOM  was first introduced in Ref.~\onlinecite{Ollitrault_2020} and was initially  applied to compute molecular excitation energies. The method was subsequently used to calculate the Green's function of a 2-site Hubbard model in Ref.~\onlinecite{Rizzo_2022} and it has seen recent extensions and applications to industrial use-cases by Asthana~\textit{et al.}~\cite{asthana2023quantum}. Below we describe the qEOM method in more detail.

In general, any excited state $|\psi\rangle_{k}$ may be expressed as an operator $\hat{O}_k$ applied to the GS $|0\rangle$ with $N$ particles
\begin{equation}
|\psi\rangle_{k} = \hat{O}_k|0\rangle. \label{eq:schr}
\end{equation}

In the qEOM method, one imposes the extra so-called annihilation condition on the operator $\hat{O}_k$
\begin{equation}
\hat{O}_k^{\dagger}|0\rangle=0. \label{eq:anhil}
\end{equation}
The time-independent Schrödinger equation for $|\psi\rangle_{k}$ can be re-expressed as
\begin{equation}
[\hat{H}, \hat{O}_k]_-|0\rangle=E_{k0}\hat{O}_k|0\rangle, \quad [\hat{A}, \hat{B}]_{\pm}:=\hat{A}\hat{B} \pm \hat{B}\hat{A},
\end{equation}
where $E_{k0}=E_{k}-E_0$ corresponds to the excitation energy. Operating on both sides of Eq.~\eqref{eq:schr} with the state $\langle 0|\hat{O}_k^{\dagger}$ and using the annihilation condition~\eqref{eq:anhil} gives
\begin{equation}
\langle 0|[\hat{O}^{\dagger},[\hat{H}, \hat{O}_k]_-]_+|0\rangle=E_{k0}\langle 0|[\hat{O}_k^{\dagger},\hat{O}_k]_+|0\rangle. \label{eq:qeom1}
\end{equation}
Finally, adding to the left and right hand side of Eq.~\eqref{eq:qeom1} their corresponding hermitian conjugate leads to\footnote{Eq.~\eqref{eq:qeom2} also follows from applying the annihilation condition to Eq.~\eqref{eq:qeom1} once again.}:
\begin{equation}
\langle 0|[\hat{O}_k^{\dagger},\hat{H}, \hat{O}_k]_+|0\rangle=E_{k0}\langle 0|[\hat{O}_k^{\dagger},\hat{O}_k]_+|0\rangle, \label{eq:qeom2}
\end{equation}
where we used the definition of the double-commutator:
\begin{equation}
[\hat{A},\hat{B},\hat{C}]_+:=\frac{1}{2}\left([\hat{A},[\hat{B},\hat{C}]_-]_+ + [[\hat{A},\hat{B}]_-,\hat{C}]_+\right).
\end{equation}
The anti-commutator in Eq.~\eqref{eq:qeom2} typically leads to cancellations in the operators (see also Sec.~\ref{sec:qeomimp} in the SI), such that the amount of operator terms is reduced compared to another similar method to compute eigenstates, called quantum subspace expansion~\cite{McClean_2017}.

The set $\{\hat{O}_k\}$ can be written in terms of a set of fermionic operators:
\begin{equation}
\hat{O}_k = \sum_j (X_k)_j\excop_j, \label{eq:basis}
\end{equation}
where the basis $\{\excop_j\}$ of excitation operators are themselves a sum of products of annihilation and creation operators $\hat{c}_k, \hat{c}_l^{\dagger}$ of the fermions in the system. A generalized eigenvalue problem (GEP) for the coefficients $(X_k)_j$ can now be derived by solving for $E_{k0}$ and imposing the stationary condition $\frac{\partial E_{k0}}{\partial (X^{\dagger}_k)_{j}}=0$ in Eq.~\eqref{eq:qeom2},
\begin{gather}
\mathbf{A}\mathbf{X}_k = E_{k0}\mathbf{B}\mathbf{X}_k, \label{eq:GEP} \\
A_{ij} = \langle 0|[\excop_i^{\dagger},\hat{H}, \excop_j]_+|0\rangle, \quad B_{ij} = \langle 0|[\excop_i^{\dagger},\excop_j]_+|0\rangle. \label{eq:GEP2}
\end{gather}
The matrix elements of \textbf{A}, \textbf{B} correspond to expectation values (EV) of the operators and can be evaluated on a QC.

For the GF, we are interested in the particle and hole states with $N+1$ and $N-1$ particles, respectively. This corresponds to the ionization potential (IP-EOM) and electron affinity equation-of-motion (EA-EOM) methods, respectively, that are well established for calculating excitation energies to ionized states with coupled-cluster ansatz using classical computation~\cite{stanton1994ipeomcc, nooijen1995eaeomcc, stanton1995steomcc}. After the entries of \textbf{A} and \textbf{B} are calculated on the QC, we solve the GEP~\eqref{eq:GEP} classically to find the coefficients $(X_k)_j$ and the excitation energies $E_{k0}$ needed for the Lehmann representation of the GF. The transition amplitudes in the numerators of the impurity GF in Eq.~\eqref{eq:lehm} are expressed in terms of EV w.r.t. the prepared GS that can be computed on the QC,
\begin{equation}
\frac{\langle n|\hat{c}_0|0\rangle}{\sqrt{\langle n|n\rangle}} = \frac{\langle 0|\hat{O}_k^{\dagger}\hat{c}_0|0\rangle}{\sqrt{\langle 0|\hat{O}_k^{\dagger}\hat{O}_k|0\rangle}} = \frac{(X_k)_l\langle 0|\excop_l^{\dagger}\hat{c}_0|0\rangle}{\sqrt{(X_k)_i(X_k)_j\langle 0|\excop_i^{\dagger}\excop_j|0\rangle}}, \label{eq:transamp}
\end{equation}
with the Einstein-summation convention used above. A graphical representation of the workflow for the impurity GF is shown in Fig.~\ref{fig:flowchartb}. Further details on our implementation of the qEOM method can be found in Sec.~\ref{sec:qeomimp} of the SI.

\subsubsection{Excitation Operators\label{sec:excop}}

The choice of excitation operators $\excop_j$ in Eq.~\eqref{eq:basis} significantly affects the solution of the GEP~\eqref{eq:GEP} and thereby the precision of the resulting GF, whenever one restricts the basis to a subset (of the complete basis of particle and hole operators). The operators can be subdivided into excitation orders singles (s), doubles (d), triples (t) etc. and take the following general form for charged excitations of particle states
\begin{equation}
\excop^{(s)}_i = \sum_j \alpha_{ij}\hat{c}_j^{\dagger}, \quad \excop^{(d)}_i = \sum_{j,k,l}\alpha_{ijkl}\hat{c}_j^{\dagger}\hat{c}_k^{\dagger}\hat{c}_l, \ldots, \label{eq:genexc}
\end{equation}
with the same so-called computational basis operators $\hat{c}, \hat{c}^{\dagger}$ that enter the AIM Hamiltonian in Eq.~\eqref{eq:HAIM}.

In this work, we use a GHF MF solution of the AIM Hamiltonian to derive our basis of excitation operators. The PySCF software package~\cite{sun_recent_2020} is used for the GHF calculation. The GHF Slater determinant (SD) with $N$ particles that approximates the exact GS of the AIM is then expressed as
\begin{equation}
|0\rangle_\text{GHF} = \hat{b}_1^{\dagger}\cdots \hat{b}_N^{\dagger}|\text{vac}\rangle, \quad \hat{b}^{\dagger}_j = \sum_k W_{jk}\hat{c}^{\dagger}_k, \label{eq:SD}
\end{equation}
where $W_{jk}$ are the molecular orbital coefficients obtained from the GHF calculation. The operators $\hat{b}_j$ are typically called Bogoliubov operators and diagonalize the corresponding MF Hamiltonian $H_\text{MF}=\sum_{i>0}\epsilon_i \hat{b}_i^{\dagger}\hat{b}_i$, with $\epsilon_i\leq\epsilon_{i+1}$ for all $i$. 

The SD's that approximate the particle and hole states of the AIM are similarly subdivided into singles (s), doubles (d), etc. and are defined by acting on the GHF GS with the appropriate amount of creation $\hat{b}^{\dagger}_i$ and annihilation $\hat{b}_i$ Bogoliubov operators for empty ($i>N$) and occupied ($i\leq N$) orbitals, respectively, of the GHF GS. 
Inspired by this, we define our set of Bogoliubov excitation operators (BEO) as
\begin{gather}
\excop^{\text{Bog}, (s)}_j = \hat{b}_j^{\dagger}, \quad \excop^{\text{Bog}, (d)}_{jkl} = \hat{b}_j^{\dagger}\hat{b}_k^{\dagger}\hat{b}_l, \ldots, \label{eq:HFph}
\end{gather}
where for each BEO, the indices of all creation operators, are either all simultaneously virtual ($j>N$) or all simultaneously occupied ($j\leq N$) orbital indices of the GHF GS. The indices of the annihilation operators in a BEO in Eq.~\eqref{eq:HFph} are all occupied orbital indices when the creation operators are virtual orbital indices and vice versa.

We note that the set of BEO's~\eqref{eq:HFph} form a basis for the excitation operators $\hat{O}_k$ that results in particle states when we apply $\hat{O}_k$ on the GS, while it also forms a basis for operators $\hat{O}_k$ that give the hole states when applying their adjoint, $\hat{O}_k^{\dagger}$ to the GS. The BEO's therefore form an independent set of fermionic operators, that make up a complete basis set for deriving all charged particle and hole eigenstates together in the qEOM method whenever the maximal excitation order is taken (see also Sec.~\ref{sec:qeomimp} of the SI). In Sec.~\ref{sec:exc_op_comparison} of the SI we show a comparison of the above BEO with another excitation operator basis choice, by computing the impurity DOS, when the GS is approximated by a VQE state.

\subsubsection{Error Mitigation \label{sec:ZNE}} 

We employ a hierarchy of error-mitigation techniques to obtain accurate results from  noisy quantum computing experiments, which constitutes the \nth{5} step shown in the flowchart of Fig.~\ref{fig:flowchartb}.

\begin{center}
 \textit{Readout Errors}    
\end{center}
First, we employ mitigation schemes to reduce read-out errors (readout error mitigation, REM). A method proposed in Ref.~\onlinecite{Bravyi2021readout} suggests the construction of a readout matrix to model the readout errors. However, the construction of such a $2^{n} \times 2^{n}$ matrix for an $n$-qubit simulation would require large computational resources as the size of the simulation increases, rendering this approach unsuitable for practical applications.

The Twirled Readout Error eXtinction (T-REx) method~\cite{vandenberg2022trex} as implemented in \verb|Qiskit|~\cite{Qiskit} is frequently used to mitigate noisy computation results for the estimate of  expectation values of observables. The T-REx method is a model-free approach which scales well compared to the previously proposed method, since the construction of the readout matrix is not required.

However, in this work we cannot directly employ T-REx (as implemented in \verb|Qiskit| \verb|Estimator|) since we need to exploit the functionality of \verb|Qiskit| \verb|Sampler| which allows to leverage  qubit-wise commutation rules~\cite{McClean2016vqe} in order to reduce the amount of Pauli string measurements (see also Sec.~\ref{sec:qeomimp} of the SI). 
Therefore, instead of T-Rex, here we employ the Matrix-free Measurement Mitigation (M3) method~\cite{nation2021scalable}. The M3 method works on a subspace of the entire readout matrix defined by the unique noisy bit strings to be corrected. The matrix elements are approximately computed using the single-qubit calibration data. The number of those unique bit strings are typically smaller than the total $2^{n}$ bit strings, a behavior which is particularly advantageous for mitigating readout errors of simulations for large systems. 

\begin{center}
 \textit{Gate Operation Errors}    
\end{center}

Second, we utilize the zero-noise extrapolation (ZNE) technique~\cite{temme2017error, kandala2018znehw} to mitigate other sources of errors such as gate operation errors (gate error mitigation, GEM). The ZNE method estimates the noiseless expectation values of observables by extrapolating the measured values at different noise levels to the zero-noise limit. The noise is systematically amplified by inserting additional gates or by stretching the duration of the microwave pulses. In this work, we amplify the noise by adding additional CNOT gates, leveraging the fact that applying $(2n+1)$ CNOT gates to the same qubit pair produces the same outcome as a single CNOT, but with larger noise. We carefully investigate the error mitigation using the ZNE method, and find that a linear extrapolation with scale factors of $n= $1.0, 1.5, and 2.0 is most suitable for our 14 qubit system, while for the 12 qubit system we use quadratic extrapolation at noise factors $n= $1.0, 2.0, and 3.0. The ZNE prototype code is used for our ZNE error mitigation~\cite{majumdar_best_2023}.

The M3 method plays an important role for small-scale problems where gate noises from the device are relatively low and the readout error is dominant, while ZNE is useful for larger simulations with high number of gate operations. In this work, we find that combining the two methods is essential to obtain meaningful expectation values for our  observables. 
However, even the combination of the two mitigation schemes (M3 and ZNE) is still not enough to provide sufficiently accurate expectation values. To improve on that, in the next section we develop an additional mitigation scheme that allows to achieve results in qualitative agreement with the reference calculations. 

\begin{center}
 \textit{ZNE-Calibration}
 \label{sec:zne-clibration}
\end{center}
\begin{figure*}[ht!]
     \begin{subfigure}[b]{0.45\textwidth}
         \centering
         \includegraphics[trim={0 0 3.9in 0},clip, width=\textwidth]{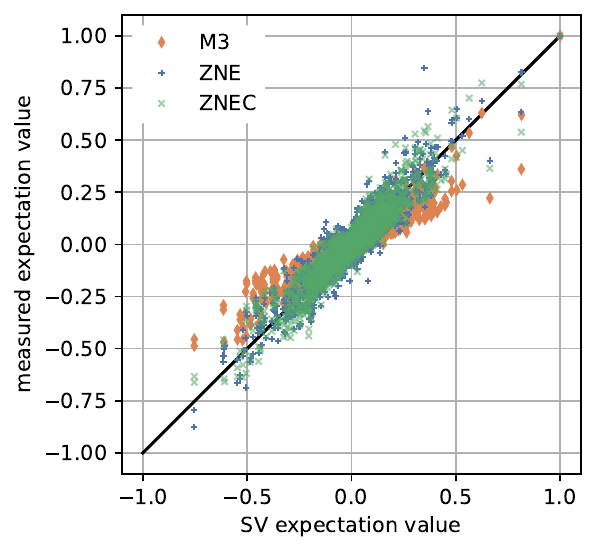}
         \caption{  }
         \label{fig:ZNE-cal-data}
     \end{subfigure}
     \begin{subfigure}[b]{0.45\textwidth}
         \centering
         \includegraphics[trim={3.9in 0 0 0},clip, width=\textwidth]{refactoring_14q.pdf}
         \caption{}
         \label{fig:ZNE-cal-fun}
     \end{subfigure}
        \caption{Demonstration of the ZNE-calibration technique for our 14 qubit experiment performed on the ``ibm\_torino'' quantum device. Subfigure (a) shows the different set of expectation values obtained using only the readout error mitigation M3 (orange, $\bar{C_j}$), using M3 and ZNE (blue, $\tilde{C_j}$), and using the ZNE-calibration technique (ZNEC, green, $C_j = f(\bar{C_j})$). The y-value of each cross represents the expectation value obtained on ``ibm\_torino'' for the calibration Pauli strings used in the ZNEC measurements of the qEOM algorithm, while the corresponding x-value stems from the simulated SV result. The data obtained with a perfect quantum computer would be placed on the diagonal black line. Subfigure (b) illustrates the calibration procedure. The red crosses show the expectation value with ZNE versus without ZNE for all calibration Pauli strings, again from the ``ibm\_torino'' device. The purple line shows the fitted function $f(x)$, used to obtain the green data in panel (a).}
        \label{fig:ZNE-calibration}
\end{figure*}

In this section, we introduce a new error mitigation strategy, which sits on top of the standard ZNE (see section~\ref{sec:ZNE}) and that we name ZNE-calibration (ZNEC). The main idea of this approach is to calibrate a function which maps results without gate-error mitigation (raw data or REM results) to the mitigated results (GEM) using ZNE  data. Compared to standard ZNE, this procedure has the advantage 
of reducing the variance of the mitigated results 
while correcting the systematic error present in the raw as well as REM results (see Fig.~\ref{fig:ZNE-cal-data}). In contrast to methods like probabilistic error cancellation or amplification (PEC~\cite{vandenBerg2023pec} or PEA,~\cite{Kim2023evidence}), ZNEC has a similar or even reduced overhead compared to standard ZNE.  
For any quantum algorithm that requires measuring a large set of Pauli strings $\hat{P_i}$ for a quantum state we perform the following steps:
\begin{enumerate}
    \item Choose a subset of calibration Pauli strings $\hat{C_j}$ of the complete set $\hat{P_i}$.
    \item Measure the Pauli strings $\hat{C_j}$ using ZNE for different noise factors. This will give the raw/REM expectation values $\bar{C_j}$ at noise factor $c=1$ and the expectation values $\tilde{C_j}$ at noise factor $c=0$ using an extrapolation method. Additionally measure all remaining Pauli strings of $\hat{P_i}$ without using ZNE to obtain their raw/REM expectation values $\bar{P_i}$. 
	\item Fit the parameter $\alpha$ of the function $f(x) = \frac{2}{\pi}\tan ^{-1}\left(\alpha\left(\tan\left(\frac{\pi}{2}x\right)\right)\right)$ to map all raw/REM results to the GEM results $\tilde{C_j} \approx f(\bar{C_j})$. The idea here is to fit only one value $\alpha$ for all Pauli strings simultaneously (see Fig.~\ref{fig:ZNE-cal-fun}). This function corrects the damping of the expectation values due to noise of the hardware, which manifests in a slope smaller than $1$ in Fig.~\ref{fig:ZNE-cal-data}.
	\item Apply the calibrated function $f(x)$ to the raw/REM results to obtain the final expectation value $P_i = f(\bar{P_i})$.
\end{enumerate}


We demonstrate the ZNEC procedure in Fig.~\ref{fig:ZNE-calibration} for a 14 qubit qEOM calculation on the ``ibm\_torino'' device. We clearly see the advantage of ZNEC over ZNE or REM (labeled with M3), without additional measurements compared to ZNE.
While ZNE is strongly affected by the statistical errors of the measurements at each noise factor $c_i$, ZNEC uses the data from 1949 operators, which is a subset of all 23354 measured operators and can be sorted into 100 commuting groups, to calibrate a functional mapping $f(x)$. This reduces the required quantum resources by a factor of approximately 3 compared to standard ZNE. For our 12 qubit results we calibrate using ZNE measurements for all Pauli strings as the total number of Pauli strings is significantly smaller than in the 14 qubit calculation. The influence of the statistical error in ZNEC is much lower than in the M3+ZNE case, and therefore we believe that this new approach can be used to significantly increase accuracy, or lower the shot budget for the evaluation of expectation values. 
Further details concerning the ZNEC scheme are discussed in Sec.~\ref{sec:ZNEC_details} of the SI.

\section{Results and Discussion}

\begin{figure}[t]
         \centering
         \includegraphics[width=0.85\columnwidth]{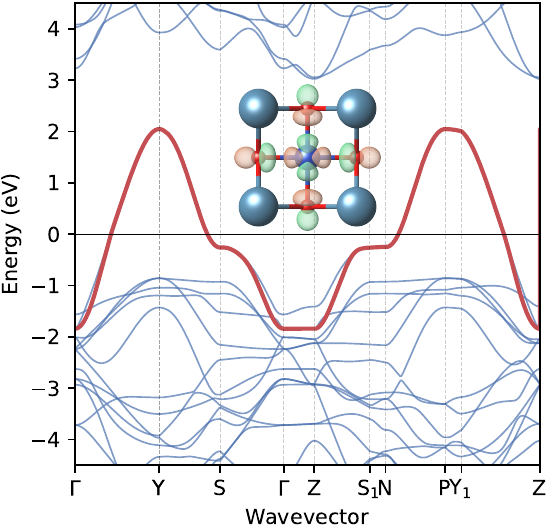}
         \caption{The electronic structure of CCOC, where the DFT Kohn-Sham bands are shown in blue (thin lines) and the interpolated Wannier band is shown in red (bold line). The Fermi level is set to zero. The inset shows a unit cell of CCOC along (001), with the large blue, small blue, and small red spheres representing the Ca, Cu, and O atoms, respectively (note that the Cl atoms are not depicted here). The positive and nevative lobes of the maximally localized Wannier function are shown as red and green isosurfaces, respectively. }
         \label{fig:CCOC}
\end{figure}

We apply the DFT+DMFT formalism with our implementation of a QC impurity solver to study the electronic structure of a high-temperature superconductor (HTS). Our material system of choice is CCOC~\cite{grande_uber_1977,yamada_single-layer_2005,baptiste_ca2cuo2cl2_2018}, for which a maximal superconducting transition temperature of around 30~K has been observed in sodium-doped samples~\cite{zhigadlo_synthesis_2008}. While this value is far below some of other HTS cuprates, e.g.,  \ce{YBa2Cu3O7}~\cite{wu_superconductivity_1987}, CCOC represents a suitable model system that exhibits the main physics of an unconventional HTS. The characteristic structural motif of copper-oxide planes within a checkerboard lattice with squares composed of \ce{O^2-} and  \ce{Cu^2-} ions at the centres is shown in the inset of Fig.~\ref{fig:CCOC}. 

We start out by constructing a low-energy model from the converged DFT electronic structure of CCOC, and transform the relevant Bloch states to localized orbitals by Wannierizing the Cu~$d_{x^2-y^2}$ state, resulting in an effective tight-binding model that reproduces well the non-interacting single-particle band crossing the Fermi level, as shown by the red interpolating line in Fig.~\ref{fig:CCOC}. The screened interaction parameter from our cRPA calculation of $U=\Re\left[U(\omega = 0)\right]=3.2$~eV is in good agreement with values for other, similar cuprate materials, e.g., of \ce{CaCuO2} in  Ref.~\onlinecite{karp_superconductivity_2022}. To obtain classical reference values and pre-converged impurity Green's functions for subsequent QC experiments, we first perform sufficient DMFT iterations with the CTHYB QMC impurity solver at finite temperatures to reach self consistency.

The converged results from the CTHYB-DMFT cycles are then fed into our fitting procedure to construct a non-interacting model with up to 6 bath sites based on the hybridization function describing the impurity problem. We then proceed by performing a self-consistent DMFT update using the ED solver at 0~K with the identical frequency mesh as in CTHYB-DMFT, then iterate by again refitting the bath parameters. A small number of merely 10 iterations suffices to reach self-consistency. The output from the final ED iteration, namely the self energy and bath parameters, serves as the input to our quantum-algorithm for computing the impurity Green's function, the detailed procedures of which will be discussed next.


\subsection{Groundstate Calculations with VQE\label{sec:VQE4GS}}

Our converged ED-DMFT AIM parameters 
are used to construct an AIM Hamiltonian in the 
qubit representation to be used in quantum hardware experiments. The first step within our quantum impurity solver is to compute an accurate, high-quality GS, which we obtain using the VQE algorithm on the noiseless SV simulator~\cite{peruzzo_variational_2014, tilly_variational_2021, cerezo_variational_2021}. To this end, we use a hardware-efficient ansatz with linear entanglement strategy~\cite{Kandala2017,ravi-22-cafqa}, and follow the procedure as described in Sec.~\ref{sec:vqe} to produce shallow quantum circuits that support the limited qubit connectivity on IBM Quantum hardware.

We quantify the quality of our GS by measuring the fidelity with respect to the exact GS from the ED results. To obtain an accurate GS, we optimize this figure of merit with respect to:  ansatz architecture (number of layers and types of rotation and entangling gates), initial state circuit (zero or GHF), initialization procedure for parameters $\boldsymbol{\theta}$ (random or identity block~\cite{Identity_Block_paper}), and random seed (used to generate initial values of parameters $\boldsymbol{\theta}$)~\cite{bonet-monroig_performance_2023}, using a grid-search based global optimization~\cite{bonet-monroig_performance_2023,selisko_extending_2023,onionvqe}. This strategy explores diverse regions of the parameter space, improving the likelihood of finding a globally optimal solution with the highest fidelity with respect to the ED GS.

In total, we sample 12 candidate ansatz architectures, encompassing all possible combinations of the following parameters:
\begin{itemize}
  \item Number of layers = 4, 6, 8,
  \item Rotation gates = RY, RY + RZ,
  \item Entangling gates = CZ, CX.
\end{itemize}
For each architecture, we explore various combinations of initial state and initial parameters $\boldsymbol{\theta}$:
\begin{itemize}
  \item Zero initial state and random initial parameters $\boldsymbol{\theta}$,
  \item GHF initial state and random initial parameters $\boldsymbol{\theta}$,
  \item GHF initial state and four variants of the identity block method~\cite{Identity_Block_paper} for initializing parameters $\boldsymbol{\theta}$:
  \begin{itemize}
    \item Initialization close to $0$~\cite{Identity} (adding a small random noise with an amplitude of 0.01),
    \item Initialization close to $\pi$~\cite{wiersema2020identity} (adding a small random noise with an amplitude of 0.01),
    \item The onion-initialization approach~\cite{onionvqe},
    \item The inverse-initialization approach, where the ansatz circuit with random initial parameters $\boldsymbol{\theta}$ is inverted and appended to the original circuit, creating an identity block and doubling the number of layers.
  \end{itemize}
\end{itemize}
Each set of initial variational parameters undergoes replication using 64 different random seeds. This results in 4608 combinations of explored parameter sets and initializations. Among all candidate states, we select an ideal GS that balances low circuit depth and high fidelity $F$~\cite{selisko_extending_2023}. We reach $F=0.959$ with 1026 single-qubit and 164 two-qubit gates for the 5-bath system, and $F=0.989$ with 252 single-qubit and 104 two-qubit gates for the 6-bath system, on the SV simulator (see Fig.~\ref{fig:ansatz} in the SI). These GS serve as the basis for the subsequent qEOM algorithm to compute the impurity GF on a quantum computer. While the VQE optimization itself is not carried out directly on the IBM Quantum hardware, we note that the complexity of the circuit structure is sufficiently low to be mapped onto a noisy quantum device. Preparing the GS state and measuring all qEOM operators, as described in the next sections, on the IBM Quantum hardware to high precision however remains a significant feat.

\subsection{Converged Selection of the Excitation Order\label{sec:EDtoSV}}

\begin{figure*}
         \centering        \includegraphics[width=0.8\textwidth]{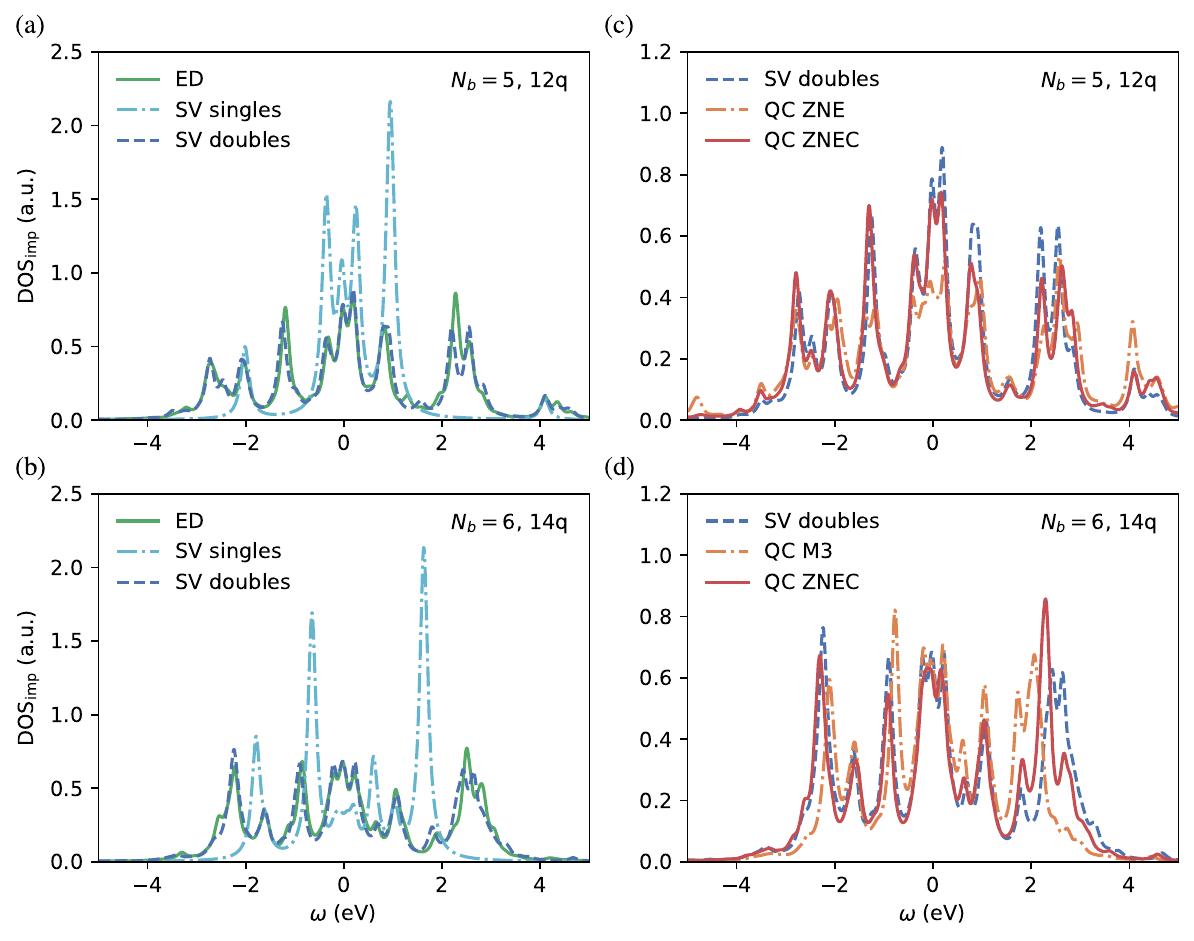}
        \caption{Impurity DOS for 5 and 6 bath sites using 12 and 14 qubits, respectively. Subfigures (a) and (b) show the results computed with the SV qEOM method using excitation operators singles, doubles, and exact diagonalization (ED) for 5 and 6 bath sites, respectively. Subfigures (c) and (d) show the results computed with the qEOM method using excitation operators up to doubles with SV and on IBM hardware.  The plots show the qEOM results with (red) and without (orange) ZNE calibration. The hardware experiments are conducted on ``ibm\_torino'' with 8192 shots to evaluate the circuits, as well as with M3 (orange in (d)), ZNE (orange in (c)), and ZNEC (red in (c) and (d)) error mitigation as explained in Sec.~\ref{sec:ZNE}. In all plots a small imaginary part of $0.1$ is added in the denominators of the GF in Eq.~\eqref{eq:lehm}.\label{fig:sd}} 
\end{figure*}

To compute the impurity GF, we employ the qEOM method as described in Sec.~\ref{sec:qeom}. Since it is challenging to \textit{a priori} determine the excitation order required to achieve a desired precision for the GF of our AIM, we first carefully compare the results for singles and doubles BEO's to determine the convergence behavior, using noiseless SV simulation for all EV. Fig.~\ref{fig:sd}~(a) and (b) compare the impurity DOS with 5 and 6 bath sites (12 and 14 qubits), respectively, taking into account single and double excitations using SV with the ED results. Here, and in all subsequent sections, we use the VQE state as described in Sec.~\ref{sec:VQE4GS} to approximate the true GS.

We observe that singles and doubles together suffice to obtain a DOS in good agreement with the ED result, both for 5 and 6 bath sites. The remaining small difference between SV and ED arise from the infidelity of the VQE state and the missing higher order excitations. For other, smaller number of bath sites we find the same behavior. Based on this observation, we elect to use singles and doubles for the calculation on noisy quantum hardware. The crucial finding that we can safely neglect all excitations exceeding doubles is the foundation for the favorable polynomial $\mathcal{O}(N^5)$ scaling (further supporting details can be found in Sec.~\ref{sec:scaling} of the SI).

\subsection{Hardware Results with ZNE-Calibration\label{sec:SVtoHW}}
Next, we compute the impurity GF on the real quantum computing QPU ``ibm\_torino'', using a hierarchy of error-mitigation techniques to improve the inherently noisy hardware results. To evaluate the circuits we set the number of shots to 8192, a value that produces sufficiently low statistical variances. As discussed in Sec.~\ref{sec:ZNE}, we then employ at the first stage the M3 method to mitigate read-out errors.

The resulting DOS of the impurity GF using up to double excitations is shown by the orange lines in Fig.~\ref{fig:sd}~(c) and (d) for 5 and 6 bath sites, respectively, using conventional ZNE and M3 data from the real hardware. Note that for the 14 qubit experiment we do not show the ZNE results due to limited compute resources. A comparison with the SV DOS (blue, dashed line) shows that the ZNE DOS in Fig.~\ref{fig:sd}~(c) and the M3 DOS in Fig.~\ref{fig:sd}~(d) exhibit the overall qualitative features of the SV results, and in particular reproduces relatively accurate excitation energies. However, the QC ZNE results produces inaccurate DOS values close to the Fermi level for the 12 qubit experiment, while the M3 DOS for the 14 qubit experiment lacks the precision required to capture the correct overlaps, and the peak positions and amplitudes diverge significantly the further away we move from the Fermi energy.

To mitigate this issue, we employ our a new error calibration scheme, ZNEC, which we introduce in Sec.~\ref{sec:zne-clibration}. The resulting calibrated DOS is shown as a red, solid line in Fig.~\ref{fig:sd}~(c) and (d) for the 12 and 14 qubit experiment, respectively (QC ZNEC). The improvement of the impurity DOS using ZNEC with respect to conventional ZNE is evident from Fig.~\ref{fig:sd}~(c) for the system with 5 bath sites, especially at the Fermi level. Compared to M3, the calibrated ZNEC in Fig.~\ref{fig:sd}~(d) effectively eliminates large portions of the remaining artifacts and corrects the majority of overlaps, in particular also those in the vicinity of the Fermi level, the most relevant portion in the energy spectrum. Residual discrepancies only remain at the peaks close to $\omega=3$~eV, an energy regime we deem not very pertinent for the subsequent steps in our workflow. The significantly improved agreement with the reference SV result demonstrates that it is of utmost importance to include our newly developed ZNEC error mitigation scheme to capture the correct physics close to the Fermi energy.


While our choice of the ZNEC calibration function is well justified in Sec.~\ref{sec:zne-clibration}, further  improvements of the error mitigation could be potentially achieved by:
\begin{itemize}
    \item adding auxiliary Pauli strings to the subset measured for the calibration. If these strings are chosen to commute with the existing measurements, there is no additional measurement effort but results in more data to train the regression model $f(x)$.
    \item improving the functional form of the regression model $f(x)$ under the constraint that it remains monotonic in the interval $[-1,1]$ and maps the values $-1$ and $+1$ to itself. In most of our calculations, the presented function $f(x)$ is able to correct a large portion of the systematic error, but a more flexible functional form, e.g., using Gaussian processes with an appropriate, monotonic and noise-aware kernel, might further improve the results.
    \item applying the same calibration scheme to other gate-error mitigation methods,  e.g., for probabilistic error cancellation or amplification~\cite{Kim2023evidence}.
\end{itemize}

\subsection{Comparison of Spectral Properties\label{sec:spectral}}
\begin{figure*}[ht!]
     \begin{subfigure}[b]{0.43\textwidth}
         \centering
         \includegraphics[height=4.7cm]{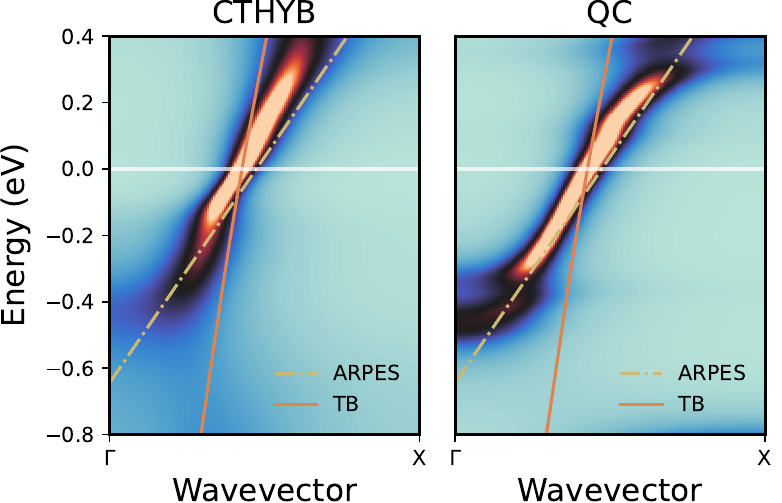}
         \caption{}
         \label{fig:Akw}
     \end{subfigure}
     \hfill
     \begin{subfigure}[b]{0.55\textwidth}
         \centering
         \includegraphics[height=4.7cm]{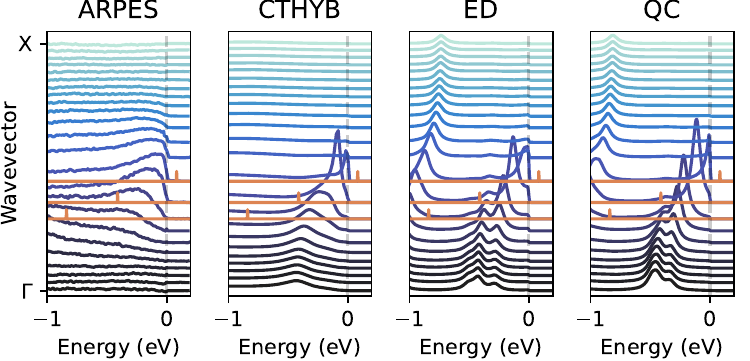}
         \caption{}
         \label{fig:ARPESvsDMFT}
     \end{subfigure}
     \hfill
        \caption{A comparison of spectral properties. Subfigure (a) shows the spectral function along $\Gamma \rightarrow X$ in the 2D Brillouin zone of the CuO-planes, with minimal and maximal values in light blue and yellow, respectively. The left panel is computed with the classical impurity solver CTHYB, while the right panel stems from our QC experiments with 14~qubits on ``ibm\_torino'' with 6 bath sites. The inscribed dashed line stems from a linear fit to the quasi particle peaks at fixed energies from experimental ARPES data (ARPES), and the solid orange line (TB) shows the non-interacting single-particle band derived from the Wannierized tight-binding model. Subfigure (b) contains a comparison of the spectral functions at discrete points along the reciprocal path $\Gamma \rightarrow X$. The first panel shows the experimental ARPES data from Ref.~\onlinecite{kohsaka_angle-resolved_2003,damascelli_angle-resolved_2003} with a subtracted background using a Gaussian process regression model. The panels denoted by CTHYB, ED, and QC show the corresponding spectral functions using the QMC impurity solver CTHYB, the ED impurity solver for the discretized representation, and the QC algorithm executed on the ``ibm\_torino'' quantum device, respectively. All panels also include the non-interacting single-particle $\delta$-peaks from the TB model in orange.}
        \label{fig:ARPES}
\end{figure*}

To obtain any physically meaningful quantities that can be compared to experimental measurements, the lattice Green's function has to be constructed from the impurity self-energy and the non-interacting single-particle dispersion. The imaginary part of this Green's function corresponds to the renormalized spectral function $A(k,\omega)$, which depends on the wavevector $k$ and energy $\omega$, a quantity that is in fact experimentally accessible through angle-resolved photo-emission spectroscopy (ARPES). However, before diving into a comparison of our results with ARPES data, we discuss the convergence behavior of our QC spectral properties with respect to classical reference calculations.

Fig.~\ref{fig:Akw} shows the spectral functions  along a predefined path in reciprocal space  $\Gamma \rightarrow X$ (where $X$ corresponds to $(\pi, \pi)$ in the 2D Brillouin zone  of the \ce{CuO} plane), using the classical CTHYB and QC impurity solver for the 6 bath sites in the left and right panel, respectively. The heat map corresponds to the spectral intensity $A(k,\omega)$, clearly showing the quasiparticle band crossing the Fermi energy at around $0.43\times (\pi, \pi)$. Within an energy window of $\pm 0.2$~eV around the Fermi level, there is good agreement between the CTHYB and QC results. However, stronger deviations are observed at energies further away from the Fermi level. These discrepancies can be attributed on one hand to the different temperatures employed for the two calculations, but predominantly to the bath discretization required for the QC solver, an artifact that would disappear with larger number of bath sites.

To quantify the differences in the spectral properties, we compare the quasi particle weight (QPW) 
\begin{equation}
 Z=\left[1-\frac{\partial\Re \Sigma(\omega)}{\partial \omega}\right]^{-1}=\left[1-\frac{\partial\Im \Sigma(i\omega_n)}{\partial \omega_n}\right]^{-1} 
 \label{eq:z}
\end{equation} 
at the Fermi level $\omega\rightarrow0$, or in Matsubara frequencies $i\omega_n\rightarrow0$~\cite{Arsenault_Benchmark_2012}. This quantity can be directly computed from the real or imaginary part of the self energy $\Sigma$, which contains the information about the renormalization of the bare electronic dispersion, and is readily available within each DMFT cycle. We obtain values of $Z_{\text{CTHYB}}=0.256$, $Z_{\text{ED}}=0.271$, and $Z_{\text{QC}}=0.265$ for the converged DMFT results using the CTHYB, ED, and QC impurity solver, respectively. While the value of $Z_{\text{ED}}$ is slightly larger than $Z_{\text{CTHYB}}$, the difference is within the range one would expect from the bath discretization. In fact, the self energy obtained from the Lehmann representation contains spurious peak structures, which influence the numerical robustness when computing the derivatives in Eq.~\eqref{eq:z}. We therefore eliminate and redistribute all peaks within $|\omega|<0.1$ as described in Sec.~\ref{sec:exact_diag}. The relative error between  $Z_{\text{ED}}$ and $Z_{\text{QC}}$ is merely 2.2\%, demonstrating the excellent agreement of our QC results with respect to the ED reference values.

Finally, we turn our attention to comparing the computational results to the experimentally measured ARPES spectra of CCOC available in the literature~\cite{kohsaka_angle-resolved_2003,damascelli_angle-resolved_2003}. The first panel in Fig.~\ref{fig:ARPESvsDMFT} plots the single-electron ARPES spectra with subtracted background signals, showing the evolution of the quasi particle peak at various wavevectors (see also Sec.~\ref{sec:backgroundarpes} of the SI). The second, third, and fourth panels contain the corresponding spectral lines computed from the converged DMFT results using the CTHYB, ED, and QC impurity solver with 6 bath sites, respectively, however convolved with a Fermi-Dirac function around the Fermi level to eliminate any spurious signals from the unoccupied states. Again, the peak evolution within an energy window of $-0.2$~eV below the Fermi level is in good agreement between the ARPES and all computed spectra.

Since the self-energy cannot be directly extracted from the ARPES spectrum to quantify the agreement, we approximate the QPW using an alternate, approximate approach by fitting a line through the maxima along the energy as a function of the wavevector $k$ close to the Fermi level (fit within $\epsilon\in[-0.4,0]$~eV and $k\in [0.25, 0.5]\times(\pi, \pi)$,  shown as a dash-dotted line in the right panel of Fig.~\ref{fig:Akw}). The ratio of its slope $m_\text{ARPES}$ and the bare electron dispersion $m_\text{TB}$, which we approximate with the TB band and is shown as a solid line in the right panel of  Fig.~\ref{fig:Akw}, provides an estimate of the QPW, $Z_{\text{ARPES}}=\frac{m_\text{ARPES}}{m_\text{TB}}$~\cite{georges_dynamical_1996}. Using this procedure, we obtain for the ARPES data a value of $Z_{\text{ARPES}}=0.274$, in excellent agreement with our values for $Z_{\text{CTHYB}}=0.256$, and the respective quantities from ED and QC. The residual discrepancies between the computed QPW and the ARPES values may be attributed to the inherent approximations involved in DMFT, such as the insufficient description of short-range correlations.





\section{Conclusions}
In this work, we present a (charge) self-consistent DFT+DMFT simulation workflow that incorporates a QC impurity solver, and demonstrate its utility by investigating the electronic structure of a prototypical HTS material on noisy quantum hardware using up to 14 qubits. The quantum algorithm to solve the underlying AIM relies on the impurity Green's function in the Lehmann representation. To obtain the excited states required for its computation, we implement the qEOM algorithm and show that a truncation after the second excitation order suffices to accurately reproduce the impurity DOS from ED results. This approximation limits the computational complexity of our algorithm to $\mathcal{O}(N^5)$, a scaling behavior that would allow applications for up to some few dozen degrees of freedom, i.e., around 10-20 bath sites for a single site AIM. Note, however, that the quantum computation can be trivially parallelized over the qEOM matrix elements, a feature that we can readily exploit on next generation quantum devices to further reduce the time to solution and hence would allow to further increase the degrees of freedom of the model.

Our experiments on IBM devices with up to 14 qubits implements the largest qEOM simulation on a real QC hardware to the best of our knowledge. This achievement is only possibly by addressing the inherent noise on current quantum computers, which poses a significant challenge as recognized in the literature~\cite{ehrlich_perspectives_2023}. Even the deployment of conventional error mitigation schemes like M3, T-Rex, and ZNE is insufficient to obtain any meaningful results. To address this issue, we develop a novel mitigation strategy, ZNEC, which relies on training a calibration function on noisy expectation values and the corresponding, gate-error-corrected counterparts to reduce systematic expectation value variances. In this work, we train an analytic calibration function that depends merely on a single fitting parameter, but more complex functional forms or an additional dependence on the circuit structure and properties of the measured observables are possible to improve the performance of the ZNEC error mitigation scheme (see also Sec.~\ref{sec:ZNEC_details} of the SI).

Devising efficient and more scalable quantum algorithms for GS calculations would represent a key step to improve our workflow and enable the solution of challenging state-of-the-art problems. Variational VQE-type algorithms~\cite{peruzzo_variational_2014, tilly_variational_2021, cerezo_variational_2021} suffer from slow convergence due to the potential presence of local minima, barren plateaus, and an overall ill-conditioned optimization behavior~\cite{mcclean_barren_2018,wang_noise-induced_2021,cerezo_variational_2021}, and are thus only suited for small system sizes~\cite{Kandala2017, peruzzo_variational_2014}. While incremental progress is constantly proposed to improve the efficiency of variational approaches~\cite{onionvqe}, more effective quantum algorithms like quantum phase estimation~\cite{kitaev_quantum_1995} and quantum imaginary time evolution (QITE)~\cite{Motta2020Qite, nishi_Qite} remain impractical on noisy quantum devices due to their associated deep quantum circuits. Variational QITE~\cite{mcardle_variational_2019} or Krylov subspace expansion~\cite{cortes_quantum_2022} might offer an affordable compromise between circuit complexity and attainable accuracy on near-term noisy quantum hardware.

Future efforts will be aimed at extending the present formalism to more complex Hamiltonians, in particular to describe multi-orbital systems as well as materials with several, symmetrically inequivalent lattice sites. Further, the implementation of the cluster extension to  DMFT~\cite{georges_dynamical_1996} would improve the description of short-range correlations, a feature required to better model the intricate phase diagrams of, e.g., HTS cuprates or nickelates. 
However, such an assessment of the phase diagram additionally requires a quantum impurity solver at finite temperatures, e.g., based on the efficient mapping of thermal states using the variational quantum thermalizer~\cite{selisko_extending_2023}.

In conclusion, the present work based on applying a scalable hybrid quantum-classical workflow to challenges beyond a simple toy model marks an important milestone en route to quantum utility for materials simulations.


\section{Acknowledgements}
We thank Sophie Beck, Alexander Hampel, and Antoine George for their valuable expert discussions on DMFT. Additionally, we gratefully acknowledge Katerina Gratsea for her insightful expert suggestions on VQE. Furthermore, we acknowledge the assistance provided by Kensuke Kurita, Kazuma Nakamura, and Takashi Koretsune with Wannier90 and RESPACK. We thank Pedro Rivero Ramirez and Nate Earnest-Noble for their support with the IBM Quantum system. JS, MA and TE thank all members of the MANIQU consortium for valuable discussions. Finally, JS thanks Ralf Drautz and Thomas Hammerschmidt for their continued mentorship. We acknowledge the use of IBM Quantum services for this work. The views expressed are those of the authors, and do not reflect the official policy or position of IBM or the IBM Quantum team.

\section{Funding}
JS, MA, and TE gratefully acknowledge support from the German Federal Ministry of Education and Research (BMBF) under project No.~13N15574. 

\section{Competing interests}
All authors declare that they have no conflicts of interest.

\section{Data Availability}
All data needed to evaluate the conclusions in the manuscript are present in the paper and/or the Supplementary Materials. Additional data that support the findings of this study are available upon reasonable request.

%


\clearpage
\onecolumngrid
\section{Supplementary Information}
\beginsupplement


\subsection{Convergence of DFT\label{sec:convergencedft}} 
\begin{figure}[h!]
         \centering      \includegraphics[width=0.85\columnwidth]{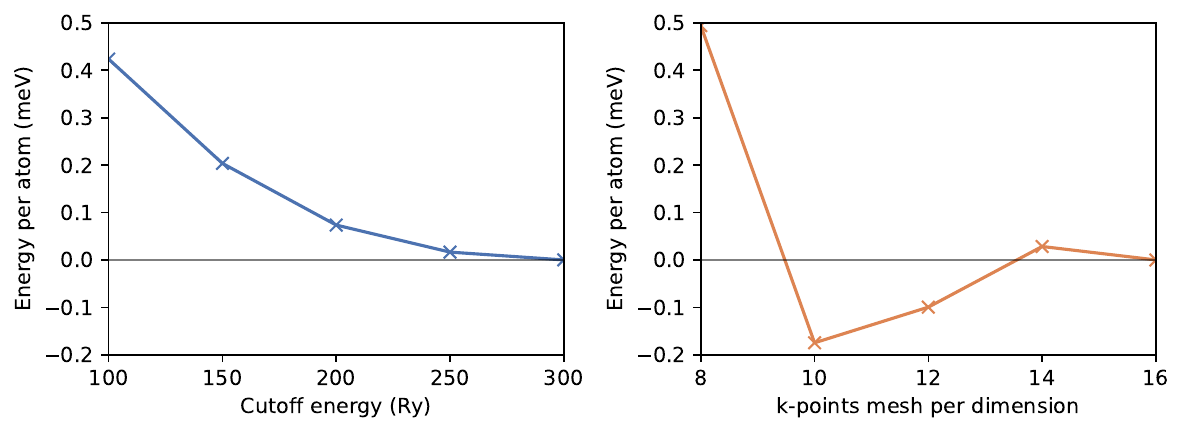}
         \caption{Convergence results of the DFT calculations with respect to the plane-wave cutoff energy and the $k$-points mesh. The total energy is shifted such that the most accurate results are set to zero.}
         \label{fig:convergencedft}
\end{figure}

We perform careful convergence tests with respect to the plane-wave cutoff energy and the $k$-points sampling of the irreducible Brillouin zone.  Fig.~\ref{fig:convergencedft} shows in the left panel the convergence of the total energy as a function of the cutoff energy, while the right panel shows the convergence behavior with respect to the number of $k$-points. Note that the x-axis in the right panel corresponds to the number of $k$-points in one dimension only, i.e., a value of 8 corresponds to a Monkhorst-Pack $k$-points mesh of $8\times8\times8$.

\subsection{Convergence of cRPA\label{sec:convergencecrpa}} 
\begin{figure}[h!]
         \centering      \includegraphics[width=0.55\columnwidth]{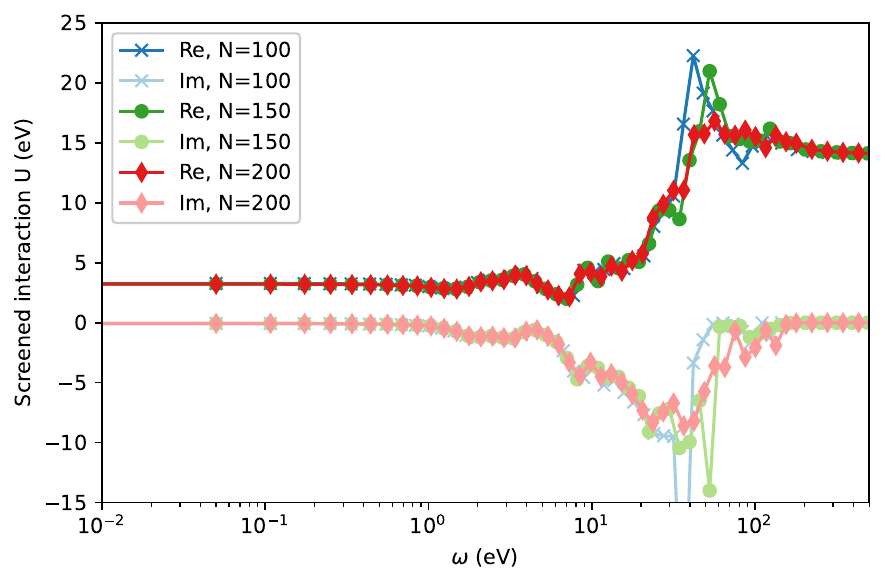}
         \caption{Convergence results of the RESPACK cRPA calculations with respect to the total number of included orbitals ($N$). The real and imaginary part of the dynamical screened interaction is plotted for three values of $N=\{100,150,200\}$.}
         \label{fig:convergenceRESPACK}
\end{figure}

We perform careful convergence tests with respect to the number of $k$-points, cutoff energy, and the number of bands included in the cRPA calculations. Fig.~\ref{fig:convergenceRESPACK} shows the real and imaginary part of the screened interaction parameter $U(\omega)$ for different number of bands $N$. Note that the static limit  $U=\Re[U(0)]$ is already well converged for $N=100$ despite larger fluctuations at finite frequencies $\omega> 1$~eV. 

\subsection{Convergence with Respect to the Number of Bath Sites\label{sec:bath_sites}}
\begin{figure}[h!]
         \centering      \includegraphics[width=0.55\columnwidth]{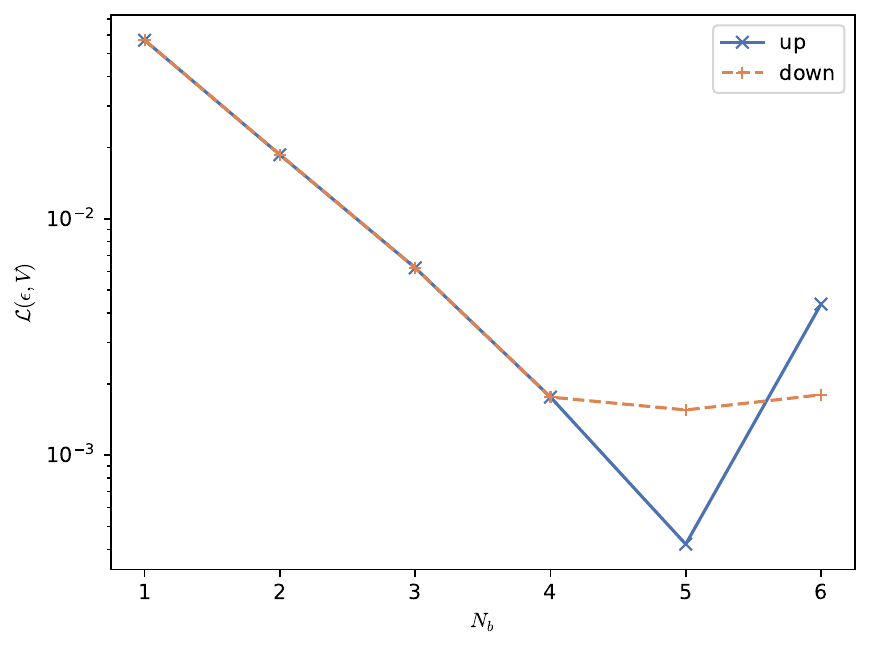}
         \caption{Convergence of the discrete bath fitting with respect to the number of bath sites. The plot shows the normalized loss function for spin up and down of the minimization problem as a function of the number of bath sites $N_b$.}
         \label{fig:convergencebath}
\end{figure}

The convergence with respect to the number of bath sites can be monitored by checking the ability of the bath to approximate the hybridization function after the final CTHYB iteration. In Fig.~\ref{fig:convergencebath} we show the loss function of the minimization in Eq.~\eqref{eq:min_hyb_bath} normalized by the number of Matsubara frequencies $N_\omega$:
\begin{equation}
    \mathcal{L}(\epsilon,V) = \sum_{\omega_n} \frac{|\Delta_\text{disc}(i\omega_n)-\Delta(i\omega_n)|}{N_\omega|\omega_n|}
\end{equation}

We clearly see that up to $4$ bath sites (10 qubits) the loss function decreases exponentially to values around $10^{-3}$. Further improvement, which is only possible for $5$ and $6$ bath sites, is challenging as the minimization process depends on more variables and the fitting gets stuck in local minima or exceeds the maximum number of iterations. As a consequence, the parameters for the up and down spin bath sites differ and we need to take the average of both for our implementation as the system is expected to have spin symmetry. Further up-scaling beyond $10$ to $14$ qubits requires an improvement of the fitting procedure by, e.g., using a continued fraction approximation, which can be performed iteratively and gives the AIM already in chain topology. The overall accuracy of our calculations is not only limited by the discretization of the bath, but also the approximate ground state on the quantum device from VQE.

\subsection{Circuits for VQE and qEOM\label{sec:ansatz}}

\begin{figure*}[h!]
     \begin{subfigure}[a]{\textwidth}
         \centering
         \includegraphics[width=0.8957\textwidth]{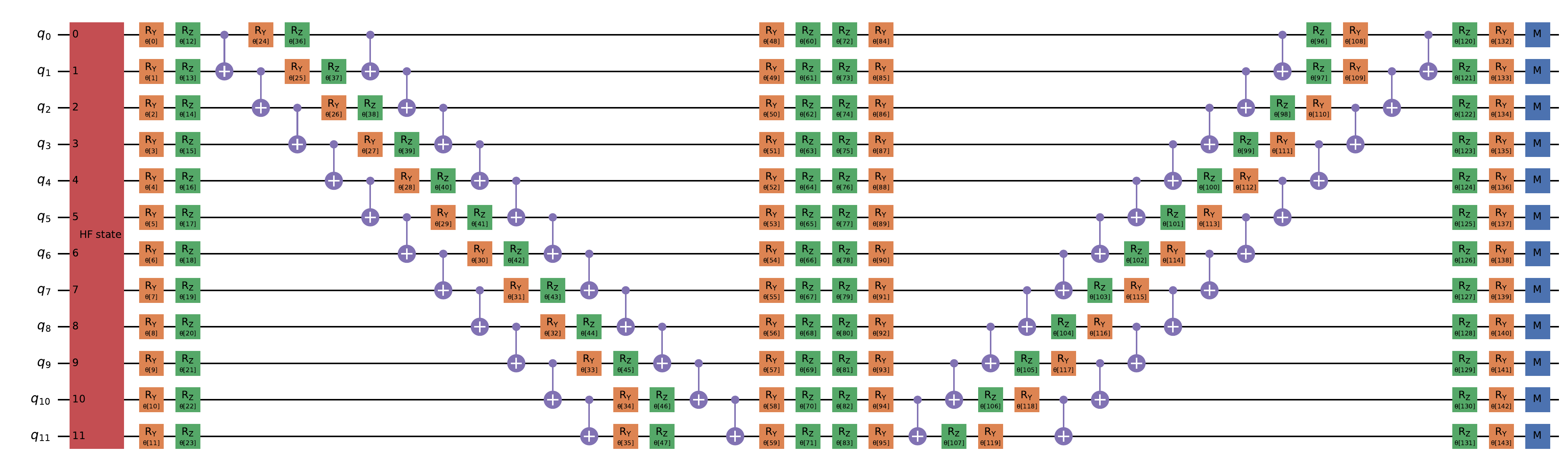}
                  \caption{ }
     \end{subfigure}
     \begin{subfigure}[b]{\textwidth}
         \centering
         \includegraphics[width=\textwidth]{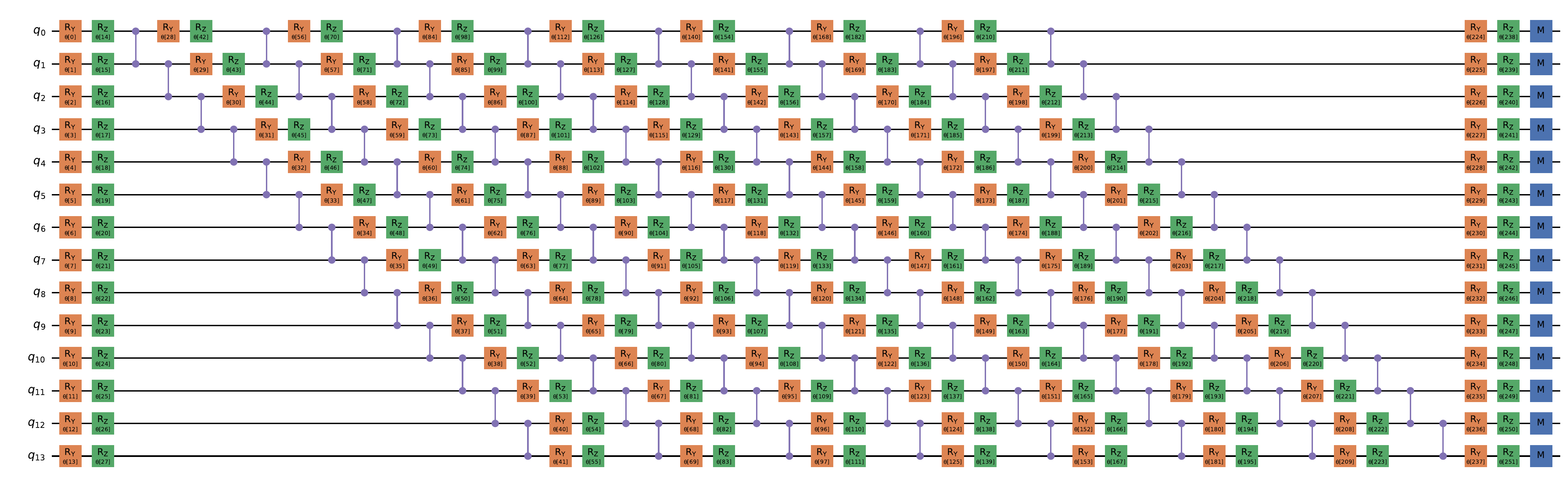}
                  \caption{ }
     \end{subfigure}
\caption{The full circuits, showing the measurement gates and the hardware-efficient ansatz with linear entanglement strategy~\cite{Kandala2017,ravi-22-cafqa} used for the construction of the GS featuring (a) 4 layers, RY and RZ rotation gates, CX entangling gates, GHF initial state, and the second half inverted for implementation of the identity block initialization for the system with 5 bath sites, and (b) 8 layers, RY and RZ rotation gates, CZ entangling gates, and zero initial state for the system with 6 bath sites. The blue boxes represent the Pauli strings of X, Y, and Z gates to perform the measurements of the Pauli strings.}
\label{fig:ansatz}
\end{figure*}

This section contains the full quantum circuits post-selected from 4608 circuits as described in Sec.~\ref{sec:VQE4GS} that are used to perform the VQE calculations and subsequently compute the matrix elements for the qEOM method. For the system with 5 bath sites, the quantum circuit has the fidelity of $0.959$, 4 layers, RY and RZ rotation gates, CX entangling gates, GHF initial state, the second half inverted for implementation of the identity block initialization, and random seed equal to $10$. For the system with 6 bath sites, the quantum circuit has the fidelity of $0.989$, 8 layers, RY and RZ rotation gates, CZ entangling gates, zero initial state, random initial parameters $\boldsymbol{\theta}$, and random seed equal to $43$. The circuits are shown in Fig.~\ref{fig:ansatz}.

\subsection{Detailed Implementation of qEOM\label{sec:qeomimp}}

For an exact GS and complete excitation basis set, an eigenstate with a positive eigenvalue $\mathcal{E}_k$ from the GEP~\eqref{eq:GEP} corresponds to a particle state $\hat{O}_k|0\rangle$ with $E_{k0}=E_k-E_0=\mathcal{E}_k$, whereas an eigenstate with a negative eigenvalue $\mathcal{E}_k$ corresponds to a hole state $\hat{O}_k^{\dagger}|0\rangle$ with $(-E_{k0})=E_0-E_k=\mathcal{E}_k$. In both cases, the denominator in the Lehmann representation of Eq.~\eqref{eq:lehm} can be written as $z - \mathcal{E}$, where $\mathcal{E}$ is an eigenvalue of the GEP~\eqref{eq:GEP}. However, in the case of an inexact GS and not complete excitation operator basis $\{\excop_k\}$ (e.g., not using the maximal excitation order) or when hardware noise is present, the eigenvalues are shifted with an error, such that an eigenstate with a positive eigenvalue does not necessarily correspond to a particle state anymore and, similarly, a negative eigenvalue state may not correspond to a hole state. This is in particular true for the small eigenvalues where sign changes might occur due to errors. Therefore, for an eigenstate with a small eigenvalue, it is unclear if it corresponds to a particle state $\hat{O}_k|0\rangle$ or a hole state $\hat{O}_k^{\dagger}|0\rangle$. 

To address the above issue, we note that in Eq.~\eqref{eq:transamp} we can replace $\hat{O}, \hat{O}^{\dagger}$ with $\hat{O} + \hat{O}^{\dagger}$, which is valid whenever the annihilation condition is satisfied and is an approximation otherwise. The same holds when we replace $\hat{c}_0$ with $\hat{c}_0^{\dagger}$ for computing the first overlap terms in Eq.~\eqref{eq:lehm}. This form of Eq.~\eqref{eq:transamp} has therefore the advantage of avoiding the problem of assigning the eigenvalues to particle ($\hat{O}_k|0\rangle$) or hole states ($\hat{O}_k^{\dagger}|0\rangle$) after solving the GEP. Our implementation is therefore more noise resilient, which is advantageous for applications using noisy quantum devices, compared to earlier implementations. In summary, the impurity GF computed after solving Eq.~\eqref{eq:GEP} with the qEOM method equals
\begin{gather}
    G_\text{imp}^{(\text{qEOM})}(z) = \sum_k \frac{|(X_k)_j\langle 0|(\excop_j + \excop_j^{\dagger})(\hat{c}_0 + \hat{c}_0^{\dagger})|0\rangle|^2}{|(X_k)_i(X_k)_j\langle 0|(\excop_i^{\dagger}+\excop_i)(\excop_j^{\dagger}+\excop_j)|0\rangle|} \nonumber \\
    \times \frac{1}{(z - E_{k0})}. \label{eq:lehmQC}
\end{gather}

The EV w.r.t. the (approximate VQE) GS that are needed for computing the matrices $\mathbf{A}, \mathbf{B}$ and those in Eq.~\eqref{eq:lehmQC} can be expressed as a sum of Pauli strings after mapping the fermionic system to qubits on the QC, using, e.g., the Jordan-Wigner transformation. After collecting the unique Pauli strings needed to compute all EV, we observe that the amount of Pauli strings increases rapidly with system size, as shown in the  Tab.~\ref{tab:scaling} of the SI. Hence, we apply the following procedures to reduce the required Pauli strings in order to efficiently execute the qEOM method on IBM hardware with up to 14 qubits:
\begin{itemize}
    \item We choose the fermionic, anti-commutator version with $[\dots]_+$ of qEOM as shown in Eq.~\eqref{eq:qeom2}. The commutator qEOM version $[\dots]_+\rightarrow [\dots]$ in Eq.~\eqref{eq:qeom2} results in an increase in the amount of Pauli strings by a factor of $\approx3-4$ for our AIM systems. This overhead arises since the
    excitation operators $\excop_j$ and AIM Hamiltonian are made up of fermionic operators $\hat{c}, \hat{c}^{\dagger}$ that naturally satisfy the usual anti-commutation relations.
    \item The spin up and down GEPs decouple, i.e. the matrix elements in Eq.~\eqref{eq:GEP2} are only non-zero whenever the two states $\hat{R}_i|0\rangle, \hat{R}_j|0\rangle$ are in the same spin-sector. We derive and solve the GEP for spin up and down excitations therefore separately, which further reduces the amount of matrix elements by a factor of 2.
    \item The total set of required Pauli strings are combined into qubit-wise commuting groups~\cite{McClean2016vqe}, using the \verb|AbelianGrouper| class in \verb|Qiskit|. The EVs of each commuting Pauli group can then be computed on the QC with a single circuit by making a corresponding single-qubit basis rotation before performing the measurement. The amount of Pauli groups are about a factor of 4 less than the total amount of Pauli strings.
    \item We only keep the explicitly real parts of the matrix elements of $\mathbf{A}, \mathbf{B}$. This is exact whenever the VQE state does not involve complex phases in its superposition and is in particular true for the exact GS. In our implementation, we could thus drop all Pauli strings with an uneven amount of $Y$-Pauli operators, which reduced the amount of measurements by about another factor of 2.
\end{itemize}

\subsection{Comparison of qEOM using Bogoliubov and computational basis excitation operators\label{sec:exc_op_comparison}}

\begin{figure}[h!]
         \centering      \includegraphics[width=0.75\columnwidth]{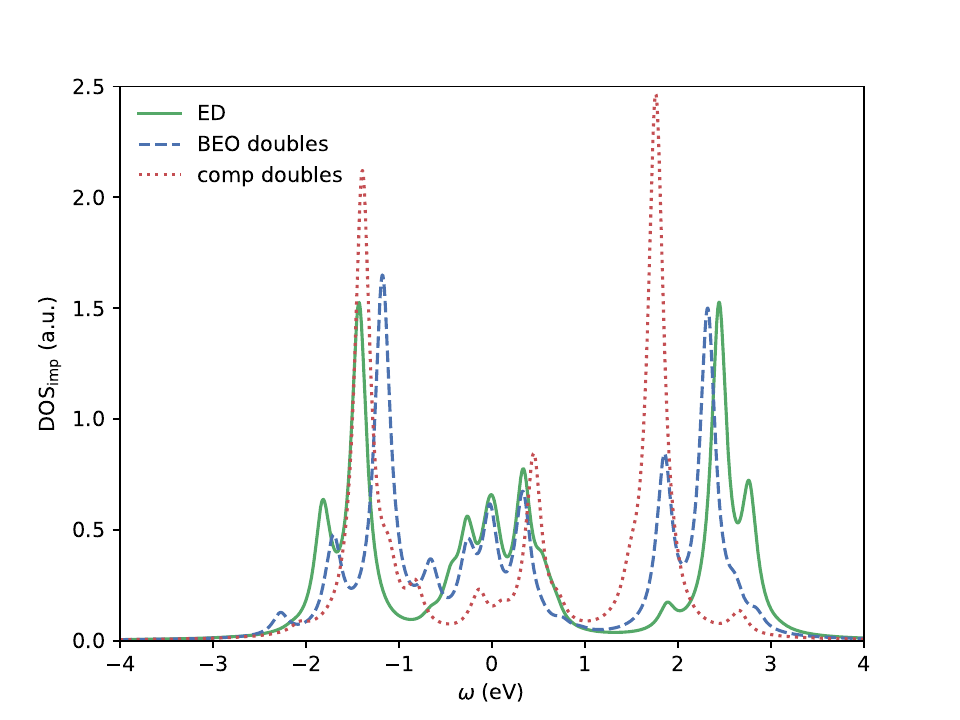}
         \caption{The impurity DOS for 2 bath sites (6 qubits) from the ED (green) and using the two types of doubles excitation operators, BEO and computational basis in blue and red, respectively, with SV qEOM as explained in the manuscript.}
         \label{fig:beo_comp}
\end{figure}

The impurity GF computed with the qEOM method depends strongly on the choice of excitation operator basis in Eq.~\eqref{eq:genexc}. There are an exponentially large (in system size) amount of alternatives to the BEO in Eq.~\eqref{eq:HFph}, which makes it difficult to find a good excitation operator basis. In our study we found empirically that for our material system, the BEO in Eq.~\eqref{eq:HFph} typically works best for producing an accurate impurity GF, when using an inexact GS prepared with the VQE algorithm. To support this observation, we show an example in this section  for the 2 bath sites case (6 qubits) and a VQE state with fidelity 0.853. Namely we compare two options for the excitation operators: the product form as in Eq.~\eqref{eq:HFph}, with either $\hat{b}$ (called the BEO in this paper) or with replacing $\hat{b}\rightarrow\hat{c}$ (called the computational basis). The VQE state has the ansatz form of the inverse-initialization approach as explained in Sec.~\ref{sec:VQE4GS} and shown in Fig.~\ref{fig:ansatz} (a), now for 6 qubits, with 2 layers and zero initial state. In Fig.~\ref{fig:beo_comp} we show the impurity DOS for the two cases, computed with SV qEOM. The BEO gives a better approximation than the computational basis one, in particular crucially around the Fermi energy. This behaviour was seen across all system sizes considered and with varying VQE fidelities. The remaining difference with the ED result arises almost exclusively from the infidelity in the VQE state.

\subsection{Detailed Analysis of the ZNE-Calibration\label{sec:ZNEC_details}}



ZNEC fits one calibration function for all Pauli strings measured on the ground state. The assumption made in the procedure, namely that the noise is independent of the operator measured on the individual qubits and even the number of identities in the Pauli string, needs some justification.

\begin{figure*}
     \begin{subfigure}[b]{0.49\textwidth}
         \centering        \includegraphics[width=0.95\columnwidth]{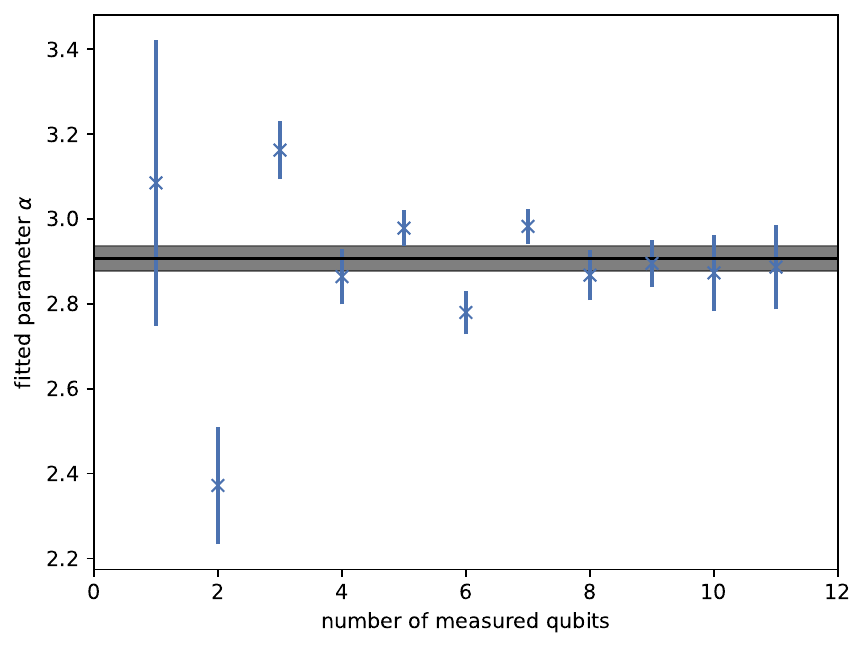}
         \caption{  }
         \label{fig:znec_nb5_number}
     \end{subfigure}
     \hfill
     \begin{subfigure}[b]{0.49\textwidth}
         \centering        \includegraphics[width=0.95\columnwidth]{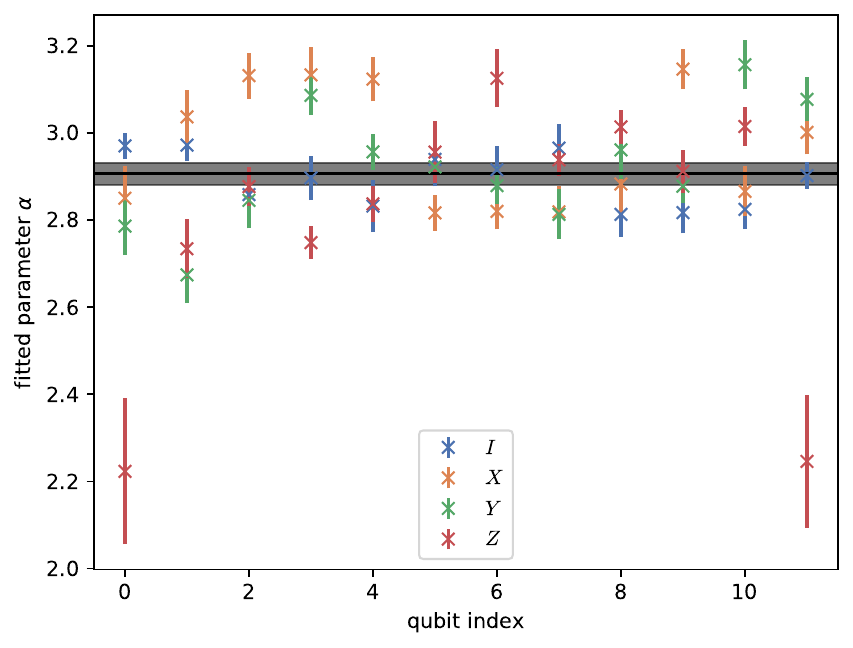}
         \caption{}
         \label{fig:znec_nb5_ops}
     \end{subfigure}
        \caption{Dependence of the fitted parameter of ZNEC on the set of calibration measurements. (a) The blue crosses show the parameter $\alpha$ fitted for Pauli strings measuring a fixed number of qubits and therefore containing a specific number of identity operators. (b) The subsets of calibration measurements are chosen to either, not measure ($I$, blue), or measure the Pauli $X$ (orange), $Y$ (green) or $Z$ (red) operator on a specific qubit (x-axis). In both subplots the black line shows the fitted parameter using all Pauli strings and has an error denoted by the grey region.} \label{fig:znec_nb5}
\end{figure*}

Fig.~\ref{fig:znec_nb5} shows numerical results which justify the validity of the above assumption for our use-case of the $12$ qubit calculation of \ce{Ca2Cl2CuO2} performed on ``ibm\_torino'', laying the basis also for $14$ qubits. In these experiments all $11259$ Pauli-strings appearing in the \textbf{A}, \textbf{B} qEOM matrices are measured using ZNE, and different subsets of operators are used to calibrate the ZNEC-function and obtain the fitted value $\alpha$. In the subfigure Fig.~\ref{fig:znec_nb5_number} the calibration subset is chosen to contain Pauli-strings which only measure a certain amount of qubits, and therefore need to have a specific number of identities. In the subfigure Fig.~\ref{fig:znec_nb5_ops} on the other hand we choose the calibration subset by selecting Pauli-strings which have a certain operator ($I$, $X$, $Y$ or $Z$) on a specific qubit, the index of which represents the x-axis. We also show an error bar which is obtained as the standard deviation of the fitted parameter $\alpha$ of $100$ random subsets of the calibration set, where the subsets contain half as many operators as the calibration set. The parameter obtained by using all Pauli-strings as the calibration set is shown as a black line and its error as the grey region (corresponds to the fit shown in Fig.~\ref{fig:ZNE-calibration} of the main manuscript). 

In Fig.~\ref{fig:znec_nb5_number} we see that the fitted parameter of the calibration function is largely independent of the number of measured qubits in the operators used to calibrate the ZNEC function. For low number of measured qubits, the data points are further away from the black line and have an increased error as there are only a few Pauli strings with $1$ to $3$ measured qubits. We also observe that measuring a Pauli $Z$ operator on the first or last qubit of the chain behaves differently to the other measurements and results in a smaller fitted value $\alpha$ in Fig.~\ref{fig:znec_nb5_ops}. 

From the results shown in this paper, we conclude that the approximation of ZNEC might not be accurate for all quantum devices, circuits, or operators, but it nevertheless significantly improves over plain ZNE results. In cases where the approximation does not hold we suggest to add more flexibility to the ZNEC function, and make it explicitly dependent on the properties (e.g., number of identities) of the Pauli strings.

\subsection{Scaling of the qEOM Algorithm\label{sec:scaling}}

The measurement of the \textbf{A}, \textbf{B} qEOM matrices constitutes the most QC-resource intensive part of the qEOM method. The amount of \textbf{A} and \textbf{B} matrix elements in the qEOM method scales quadratically with the amount of excitation operators, which itself scales exponentially with the system size. For this reason one needs to limit the excitation order for the computation and in this work we took singles and doubles into account. An insightful metric for assessing the scaling of the quantum part of the qEOM algorithm is the quantity of Pauli strings that must be measured for the \textbf{A} and \textbf{B} matrices\footnote{The Pauli-strings appearing in the transition probabilities for the impurity GF are only computed after solving the GEP and are therefore dependent on the approximate GS prepared on the QC. In our experience, the amount of Pauli strings needed for the GF is about an order of magnitude less than those in the \textbf{A}, \textbf{B} matrices.}. 

\begin{table}[h!]
 \caption{The total number of Pauli strings appearing in the \textbf{A}, \textbf{B} matrices with the singles and doubles excitation operators for \ce{Ca2CuO2Cl2}.}\label{tab:scaling}
\begin{ruledtabular} 
  \begin{tabular}{lcccccc}
    Qubits & 4 & 6 & 8 & 10 & 12 & 14 \\ \hline
    Paulis & 39 & 326 & 1497 & 4632 & 11259 & 23354 \\
  \end{tabular}
 
  \end{ruledtabular} 
\end{table}

Tab.~\ref{tab:scaling} shows the number of Pauli-strings needed for performing the qEOM method with singles and doubles excitation operators for the material \ce{Ca2CuO2Cl2}. From the table it follows that the Pauli strings count increases polynomially with a power of 5. Scaling the method to 20 qubits therefore requires about $\sim 120000$ Pauli string measurements. To perform the large number of measurements, further simplifications beyond those conducted in this paper are necessary.
One such extension is to study the relevance of the excitation operators by considering their contributions to a GHF state for the GS, but we leave this for future study.

\subsection{Background Subtraction of ARPES Data\label{sec:backgroundarpes}} 
\begin{figure}[h!]
         \centering      \includegraphics[width=0.65\columnwidth]{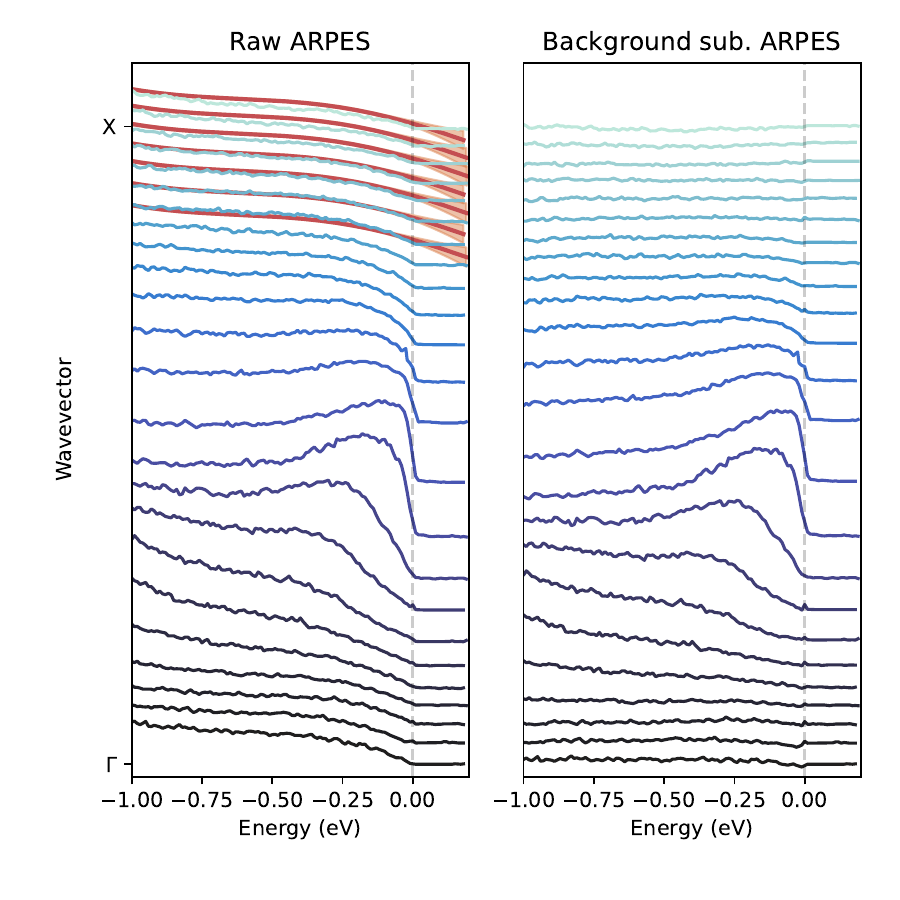}
         \caption{The raw and background subtracted ARPES data in the left and right panel, respectively. The red lines indicate the background model, together with the 99\% confidence interval shown with the orange shaded domain.}
         \label{fig:arpesbackground}
\end{figure}

We extract the ARPES results by digitizing the plots from Fig. 20b in Ref.~\onlinecite{damascelli_angle-resolved_2003}. Before comparing with the computed results, we subtract the background based on fitting a flexible function to the signal within the range of wavevectors and energies where no spectroscopic data due to single electron scattering is expected. For this purpose we fit a Gaussian process regression model to the 7 spectra closest to $X$ with the energy range up to the Fermi level. A radial basis function (RBF) kernel with an initial length scale of 1~eV together with a small noise fraction of 0.01 is used to train the model, and the hyperparameters are optimized by maximizing the log-marginal-likelihood. Fig.~\ref{fig:arpesbackground} shows the raw and background subtracted ARPES data, together with the Gaussian process background model.

\end{document}